\documentclass[aps,twocolumn,prl,floats,superscriptaddress,amsmath,amssymb,floatfix]{revtex4}
\RequirePackage{xspace} 

\usepackage{graphicx}
\usepackage{color}
\usepackage{mathrsfs}
\usepackage{amssymb}
\usepackage{amsmath}
\usepackage{epsfig}
\usepackage{url}

\usepackage{epsfig}
\usepackage{graphicx}
\usepackage{color}
\usepackage{mathrsfs}
\usepackage{amssymb}
\usepackage{amsmath}
\usepackage{epsfig}
\usepackage{url}

\def\be {\begin{equation}}
\def\ee {\end{equation}}
\def\bea {\begin{eqnarray}}
\def\eea {\end{eqnarray}}
\def\bc {\begin{center}}
\def\ec {\end{center}}
\def\n {\nonumber}
\def\sss{\scriptscriptstyle}
\def\bfl{\begin{flushleft}}
\def\efl{\end{flushleft}}
\def\bfr{\begin{flushright}}
\def\efr{\end{flushright}}

\begin{document}

\title{ A complete determination of New Physics parameters in leptonic
  decays of $\mathbf{B_{s}^0}$}

\author{Basudha Misra~\footnote{E-mail: basudha@bits-hyderabad.ac.in}}
\affiliation{BITS-Pilani, Hyderabad Campus, India}
\author{Jyoti Prasad Saha~\footnote{E-mail: jyotiprasadsaha@gmail.com}}
\affiliation{Kalyani University, India}

\begin{abstract}
Recently LHCb and CMS have measured the branching ratio of the rare decay of $B_s^0\rightarrow \mu^+ \mu^- $ which is at par with the predicted standard model (SM) value. This result does not predict the absence of new physics (NP), rather it confines the NP parameter-space in a more stringent fashion. In this paper we have used the general loop level Hamiltonian to constrain the parameter space for the NP couplings for $B_s^0\rightarrow \mu^+ \mu^- $ from the experimental branching ratio. Using the available parameter space for these couplings, we have explored the angular analysis of the cascade decay of $B^0_{s}\to \tau^+\tau^- \to (\rho^+\bar{\nu}_\tau)_{\tau^+}(\pi^{-}\nu_{\tau})_{\tau^-} \to ((\pi^+ \pi^0)_{\rho^+}\bar{\nu}_\tau)_{\tau^+}(\pi^{-}\nu_{\tau})_{\tau^-}$. This analysis shows presence of NP can be identified by the angular analysis and also shows how to isolate the NP contribution from the SM. We believe that in future with the reduced error in the branching ratio measurement and better $\tau$ cone reconstruction technique we will be able to probe NP signal using this angular analysis. 
\end{abstract}

\maketitle
\bfl
\efl

\section{I. Introduction}
\label{sec:Intro}

The rare decays of neutral $B_s^0$-mesons into 2-body leptonic modes, $B_s^0 \to l^+l^-$ for $l=e,\mu,\tau$ can happen in Standard Model(SM) only through flavour changing neutral current(FCNC) processes. They cannot occur through tree level processes and can be mediated by only electro-weak box and Z penguin diagrams and are extremely rare due to helicity suppression by $(m_l/M)^2$ factor where $m_l$ and $M$ are the masses of the lepton $l=e,\mu,\tau$ and $B_s^0$ respectively. These decays are further suppressed by $(f_{B_s^0}/M)^2$ due to an internal quark annihilation within the $B_s^0$ meson, where $f_{B_s^0}$ is the decay constant of the $B_s^0$ meson. In SM, the main uncertainty comes due to the partial knowledge of the decay constant and CKM matrix elements involved in these branching ratios($\mathcal{BR}$). Though recently using different lattice QCD methods, error has been reduced in the estimation of these decay constants \cite{decay-const}. In SM, \cite{sm-br},
\bea
\label{eq:sm-BR}
\mathcal{BR}(B^0_s\to e^+e^-) &\approx& \mathcal{O}(10^{-13}), \n\\
\mathcal{BR}(B^0_s\to \mu^+\mu^-)&=&(3.2\pm0.2)\times10^{-9}.\\
\mathcal{BR}(B^0_s\to \tau^+\tau^-)&\approx& \mathcal{O}(10^{-6}- 10^{-7}). \n
\eea
Due to smallness of the $\mathcal{BR}(B^0_s\to e^+e^-)$, verifying it is beyond the scope of present experimental limit. The $\mathcal{BR}(B^0_s\to \tau^+\tau^-)$ is the largest of the helicity suppressed purely leptonic decays, since the amplitude is proportional to the mass of the lepton $m_l$. However, the $\tau$ leptons are difficult to detect making the detection of this mode uniquely challenging. The complications for this mode arise not only from the combinatorial background but also from the fact that at least the two $\nu_\tau$ from the $\tau$ lepton decays are undetected and hence the two $\tau$ leptons cannot be fully reconstructed. At present we do not have exact experimental numbers for most of these $\mathcal{BR}$s, but recently LHCb collaboration \cite{exp-b-mumu} has given experimental data on
\bea
\label{eq:exp-BRmu1}
\mathcal{BR}(B^0_{s}\to \mu^+\mu^-) &=& 2.9^{+1.1}_{-1.0} \times10^{-9}
\eea
and CMS has given \cite{exp-b-mumu-cms}
\bea
\label{eq:exp-BRmu2}
\mathcal{BR}(B^0_{s}\to \mu^+\mu^-) &=& 3.0^{+1.0}_{-0.9} \times10^{-9}.
\eea

In this paper we mainly concentrate on $(B^0_{s}\to l^+l^-)$ channel, except $l=e$. We consider that NP contribution in $B_{s}^0 \to l^+l^-$ can occur only through loop level process like its SM counterpart. With this assumption, the effective Hamiltonian for the process $B_{s}^0 \to l^+l^-$ can be written in terms of three distinct Dirac bilinear contributions as (almost same as \cite{Hamiltonian})
\begin{eqnarray}
\label{eq:hamiltonian}
\mathcal{H}_{eff}=-(G_F\alpha/2\sqrt{2}\pi)(V^{*}_{ts} V_{tb})\{
C_{AA}(\bar{s}\gamma_{\mu}\gamma_5 b)(\bar{l}\gamma_{\mu} \gamma_5
l)\nonumber\\ 
 +C_{PS}^l((\bar{s}\gamma_5 b)(\bar{l} l)+C_{PP}^l(\bar{s}\gamma_5
 b)(\bar{l} \gamma_5 l)\} 
\end{eqnarray}
and the matrix element for the process can be written as \cite{matelement}:
\begin{equation}
\label{eq:matelment}
\mathcal{M} = if_{B_s}(G_F\alpha/2\sqrt{2}\pi)(V^{*}_{ts}
V_{tb})[A_l(\bar{l} \gamma_5 l)+B_l(\bar{l} l)], 
\end{equation}
where,
\begin{eqnarray}
\label{eq:AB}
A_l &=& 2m_l C_{AA}- \frac{M^2}{m_b + m_s }C_{PP}^l, \\
B_l &=& - \frac{M^2}{m_b + m_s }C_{PS}^l. \nonumber
\end{eqnarray}

$C_{AA}$ is the axial vector - axial vector type interaction term. It is present in the SM and in general, this type of coupling may be present in NP too. The $m_l$ factor before $C_{AA}$ comes due to helicity suppression. We divide $C_{AA}$ into two terms $C_{AA}^{SM}$ and $C_{AA}^{NP}$. $C_{AA}^{SM}$ is solely the SM part and $C_{AA}^{NP}$ is coming due to the NP contribution and we call it as $NP_1$. $C_{PP}^l$ and $C_{PS}^l$ are the pseudo scalar - pseudo scalar and pseudo scalar - scalar type interaction terms respectively. $C_{PP}^l$ can have a possible SM contribution from a neutral Goldstone boson penguin diagram \cite{gbpenguin} and $C_{PS}^l$ can have a possible SM contribution from SM Higgs penguin diagram \cite{higgspenguin}. But these contributions from SM higgs and neutral Goldstone bosons to the amplitude are further suppressed by $\frac{m_b^2}{M_W^2}$ compared to the dominant contribution. That is why the SM contributions in $C_{PP}^l$ and $C_{PS}^l$ can be ignored and it can be considered that they are coming from purely NP contribution. For our discussion we distinguish these two terms as $NP_2$ and $NP_3$ respectively. In this Hamiltonian as both NP and SM are appearing at loop level, we have separated out $\alpha$ and CKM elements from the $C$ couplings unlike \cite{Hamiltonian}. \\

In SM, $C_{AA}$ coupling is identical for the process $B^0_s\to l^+l^-$ for $l=\mu$ and $l=\tau$, whereas in existing NP models like MSSM, 2HDM, top-color assisted LHT models, it has been shown that $C_{PP}^l$ and $C_{PS}^l$ get $m_l$ suppression factor \cite{mlsup} due to the involvement of various NP particles in box, penguin and fermion self-energy diagrams. But no such effect of $m_l$ suppression factor has been predicted for $C_{AA}^{NP}$ coupling in existing literature. In our paper, we consider $C_{AA}^{SM}$ and $C_{AA}^{NP}$ are same, whereas $C_{PP}^l$ and $C_{PS}^l$ have a $m_l$ suppression factor for $l=\mu$ and $l=\tau$ in $\mathcal{BR}(B^0_s\to l^+l^-)$. With this assumption, it can be seen from Eq.~(\ref{eq:AB}) that both $B_l$ and $A_l$ will differ for different $\mathcal{BR}(B^0_s\to l^+l^-)$ due to the involvement of different lepton masses.\\

LHCb and CMS data of the $\mathcal{BR}(B^0_{s}\to \mu^+\mu^-)$ is in the same ballpark of the SM, but it is not ruling out the possibility of the presence of new Physics coming from any one of the three different type of NP interaction terms. We have to wait till the experimental error reduces to have any conclusive remark on the presence of new physics in this leptonic decay mode. There is a possibility that all of $NP_1$, $NP_2$ and $NP_3$ terms are present but they are cancelling each other in such a fashion that at the end of the day the SM term gives the main contribution in the branching ratio. Another possibility is that any one of the NP and SM terms are mutually cancelling each other and contribution from rest of the two NP terms satisfies the experimental data. Our main intention is to construct such observables which can separate out the effect of the presence of the NP contribution from its SM counterpart or at least identify the prominent presence of NP. \\

Considering the general Hamiltonian, it has been shown that the $C_{AA}$ term is nothing but the Willson coefficient $C_{10}$ \cite{wc}. We are not taking the value of $C_{AA}$ exactly same as $C_{10}$ as there is some uncertainty. We constrain the parameter space for $C_{AA}^{SM}$ from the SM estimated value of $\mathcal{BR}(B^0_{s}\to \mu^+\mu^-)$ and then using this constraint we figure out the allowed parameter space for the modulus of $C_{AA}^{NP}$, $C_{PP}^{\mu}$ and $C_{PS}^{\mu}$ from the experimental data of $\mathcal{BR}(B^0_{s}\to \mu^+\mu^-)$. From these constraints we can figure out the sizes for the modulus of $C_{PP}^{\tau}$ and $C_{PS}^{\tau}$, whereas we cannot obtain any bound on the phase factor of these couplings. We choose that the phases can vary from $0-\pi$ for the $C$ couplings involved in $B^0_{s}\to \tau^+\tau^-$ as in general these couplings can be complex. Using these constraints we figure out the sizes of $A_{\tau}$ and $B_{\tau}$. With these allowed sets of $A_{\tau}$ and $B_{\tau}$ we present an angular analysis of the cascade decay of $B^0_{s}\to \tau^+\tau^- \to (\rho^+\bar{\nu}_\tau)_{\tau^+}(\pi^{-}\nu_{\tau})_{\tau^-} \to ((\pi^+ \pi^0)_{\rho^+}\bar{\nu}_\tau)_{\tau^+}(\pi^{-}\nu_{\tau})_{\tau^-}$ in a model independent fashion to check whether we can separate out $B_{\tau}$ (which is purely NP contribution) from $A_{\tau}$ or we can establish a measurable difference between SM and NP effect due to $A_{\tau}$ itself. Isolation of $B_{\tau}$ will help us further to constrain the scalar sector. Here we would like to mention that it has been shown in literature that NP contribution in $\mathcal{BR}(B^0_s\to \tau^+ \tau^-)$ can be much larger, ~ 3\% - 10\% \cite{Hamiltonian},\cite{damol} which will open up many more ways to find NP signature. \\

In Section II we constrain the parameter space for $C_{AA}^{SM}$(SM), $C_{AA}^{NP}$($NP_1$), $C_{PP}^{\mu}$($NP_2$) and $C_{PS}^{\mu}$($NP_3$), using the experimental branching ratio for $B^0_{s}\to \mu^+\mu^-$. For this purpose we quote the relevant formulae and present the necessary numerical input. Section III.A deals with the cascade decays of $B^0_{s}\to \tau^+\tau^-$. The choice of reference frame for this cascade decay is discussed in detail. The angular analysis for the cascade decay is discussed in Section III.B. Various observables which help to establish the fact that the presence of NP makes a measurable difference from SM are discussed in this section too. We conclude and summarize in Section IV. In Appendix, we present the matrix element square for the cascade decay of $B_s^0$.

\section{ II. Constraints from $B^0_{s}\to \mu^+\mu^-$}
\label{sec:Bmumu}


In this section first we figure out the constraints on $C_{AA}^{SM}$ from the theoretical branching ratio of $B^0_{s}\to \mu^+\mu^-$, considering all the NP contribution as zero. Now in SM, 

\be
\begin{split}
\label{eq:smbr}
\mathcal{BR}(B^0_s\to \mu^+\mu^-) = \frac{M m_{\mu}^2}{32\pi^3} & f_{B_s^0}^2 G_F^2 \alpha^2 |V_{ts}|^2|V_{tb}|^2\tau_{B_s^0}\\
 & \sqrt{1-4\frac{m_{\mu}^2}{M^2}}|C_{AA}|^2
\end{split}
\ee
Where $\tau_{B_s^0}$ is the lifetime of $B_s^0$. Following the general Hamiltonian from Eq.~(\ref{eq:hamiltonian}) and considering that all the couplings are present, the branching ratio can be written as:
\be
\begin{split}
\label{eq:npbr}
\mathcal{BR}(B^0_s\to \mu^+\mu^-) =  \mathcal{C} \Big[c_{1\mu}^2| &  C_{AA}|^2+c_2^2|C_{PP}^{\mu}|^2+c_2^2 c_3^2|C_{PS}^{\mu}|^2 \\
&  -2c_{1\mu} c_2\mathcal{R}eal(C_{AA}C_{PP}^{\mu *})\Big]  
\end{split}
\ee
where,
\bea
\label{eq:C}
|C_{AA}|&=& \sqrt(|C_{AA}^{SM}|^2 +|C_{AA}^{NP}|^2+2\mathcal{R}eal(C_{AA}^{SM} C_{AA}^{NP*}))\n\\
\mathcal{C} &=& \frac{M }{128\pi^3} f_{B_s^0}^2 G_F^2 \alpha^2 |V_{ts}|^2|V_{tb}|^2\tau_{B_s^0}c_3 \n 
\eea
\bea
\label{eq:c1c2c3}
c_{1\mu} &=& 2m_{\mu} \n \\
c_2 &=& \frac{M^2}{m_b + m_s} \n \\
c_3 &=& \sqrt{1-4\frac{m_{\mu}^2}{M^2}}\n
\eea

\begin{table}
 \begin{tabular}{| c | c | c |}
\hline
{\bf Observables} & {\bf Value} & {\bf Reference }\\
\hline
$M$ & $5366.77 \pm 0.24$ MeV & \cite{pdg}\\
\hline
$m_{\mu}$ & $105.66$ MeV& \cite{pdg}\\
\hline
$m_{\tau}$ & $1776.82 \pm 0.16$ MeV & \cite{pdg}\\
\hline
$m_{b}$ & $4660 \pm 30$ MeV & \cite{pdg}\\
\hline
$m_{s}$ & $95 \pm 5$ MeV & \cite{pdg}\\
\hline
$f_{B_s^0}$ & $224 \pm 5$ MeV & \cite{fbs}\\
\hline
$G_F$ & $1.166 \times 10^{-11}  {\rm{MeV}}^{-2} $ & \cite{pdg}\\
\hline
$V_{ts}$ & $0.04073^{+0.0012}_{-0.0018}$  & \cite{ckmfitter}\\
&$(\pm 3\sigma)$&\\
\hline
$V_{tb}$ & $0.999132^{+0.000076}_{-0.000052}$  & \cite{ckmfitter}\\
&$(\pm 3\sigma)$&\\
\hline
$\tau_{B_s^0}$ & $(1.516  \pm 0.011)\times 10^{-12}$ s & \cite{pdg}\\

\hline
\end{tabular}
  \caption{Numerical inputs of Eq.~(\ref{eq:smbr})} 
  \label{tab:bsmumu}
\end{table}

\begin{table*}[thb]
\bc
\begin{tabular}{| c | c | c | c | c | c |}
\hline
 {\bf Case} & {\bf Allowed situation } & {\bf Range of  } & {\bf $\Gamma \times 10^{22} $} & {\bf  $|B_\tau|$} & {\bf $\Gamma  \times 10^{22}$ }\\
&&$|A_\tau|$(GeV)& when  & (GeV)& including  \\
&&&$|B_\tau|=0$&(Max.)&$|B_\tau|$(Max)\\
\hline
I & SM & $19-21.5$ & $ 52.7 - 67.5$ & 0 & $ 52.7 - 67.5$\\
\hline
II & SM+$NP_1$ & $0.65-64.6$ & $ 0.0617 - 615$ & 0 & $ 0.0617 - 615$ \\
\hline
III & SM+$NP_2$ &  $0.3-63.8$ &  $ 0.0131 - 594$ & 0 & $ 0.0131 - 594$\\
\hline
IV & SM+$NP_3$ &  $19-21.5$ & $ 52.7 - 67.5 $ & $11.4$ & $ 103 - 121$ \\
\hline
V & SM+$NP_1$+$NP_2$ &  $0.35-83.3$ & $ 0.0159 - 1010$ & 0 & $ 0.0159 - 1010$  \\
\hline
VI & SM+$NP_1$+$NP_3$ & $1.5-63.3$ & $ 0.328 - 585$ & $22.8$ & $ 56.2 - 670$\\
\hline
VII & SM+$NP_2$+$NP_3$ & $0.2-62.7$ & $ 0.00584 - 574$ & $22.7$ & $ 54.7 - 658$\\
\hline
VIII & SM+$NP_1$+$NP_2$+$NP_3$ & $0.95-81.6$ & $ 0.131 - 972$ & $22.8$ & $ 55.7 - 1070$ \\
\hline
\end{tabular}
  \caption{Allowed range of $|A_\tau|$ and maximum $|B_\tau|$ obtained from the experimental branching ratio are mentioned in third and fifth column. The range of decay width of the cascade decay $B^0_s\to\tau^{+}\tau^{-}\to(\rho^+\bar{\nu}_\tau)_{\tau^+}(\pi^{-}\nu_{\tau})_{\tau^-}\to ((\pi^+\pi^0_{\rho^+}+\bar{\nu}_\tau)_{\tau^+}(\pi^{-}\nu_{\tau})_{\tau^-}$ for our Analysis-I for various ranges of $|A_\tau|$ in the absence of $B_\tau$ and in the presence of maximum $|B_\tau|$ are mentioned in fourth and sixth column respectively.} 
  \label{tab:decaywidth}
\ec
  \end{table*}

The modulus of the couplings $A_\mu$, $B_\mu$, $A_\tau$ and $B_\tau$ has the following relations with the $C$ couplings:


\be
\begin{split}
\label{eq:modAtau}
|A_{\mu(\tau)}| = \Big[  & c_{1\mu(\tau)}^2  |C_{AA}^{SM}|^2 + c_{1\mu(\tau)}^2|C_{AA}^{NP}|^2 + c_2^2|C_{PP}^{\mu(\tau)}|^2 \\
                      & + 2c_{1\mu(\tau)}^2|C_{AA}^{SM}||C_{AA}^{NP}|{\rm cos}\, \phi_{AN} \\
                      & - 2c_{1\mu(\tau)} c_2|C_{AA}^{SM}||C_{PP}^{\mu(\tau)}|{\rm cos}\,\phi_{PP}^{\mu(\tau)}  \\
                      & - 2c_{1\mu(\tau)} c_2|C_{AA}^{NP}||C_{PP}^{\mu(\tau)}|{\rm cos}\,(\phi_{AN} -\phi_{PP}^{\mu(\tau)} ) \Big]^\frac{1}{2}
\end{split}
\ee
and 
\bea
\label{eq:Bmutau}
|B_{\mu(\tau)}| &=& c_2 |C_{PS}^{\mu(\tau)}|
\eea

where $c_{1\tau} = 2m_\tau$, $\phi_{AN},\,\phi_{PP}^\mu, \, \phi_{PP}^\tau $ are the phases of $C_{AA}^{NP}, \, C_{PP}^\mu, \, C_{PP}^\tau$ respectively. We choose the phase factor of $C_{AA}^{SM}$ as zero without any loss of generality as only the relative phase between two couplings is important. Rest of the $C$ couplings have the following relations for these decay channels: \\ 

\bea
\label{eq:Cmutau}
|C_{PP}^{\mu}|&=& m_{\mu}|C_{PP}|\n\\
|C_{PP}^{\tau}|&=& m_{\tau}|C_{PP}|=\frac{m_{\tau}}{m_{\mu}}|C_{PP}^{\mu}|\\
|C_{PS}^{\mu}|&=& m_{\mu}|C_{PS}|\n\\
|C_{PS}^{\tau}|&=& m_{\tau}|C_{PS}|=\frac{m_{\tau}}{m_{\mu}}|C_{PS}^{\mu}|\n
\eea

Here $|C_{PP}|$ is same for both $\mu$ and $\tau$ and $|C_{PS}|$ is also same for both $\mu$ and $\tau$. The magnitude of these couplings have been expressed in some particular NP models \cite{mlsup}. In our paper first we obtain bounds on the modulus of the $C$ couplings involved in $B^0_{s}\to \mu^+\mu^-$ using the experimental branching ratio data from Eq.~(\ref{eq:exp-BRmu1}). Table.~(\ref{tab:bsmumu}) contains the data used for this calculation. Experimental error has been included in each numerical input used in the branching ratio. With these bounds we restrict the modulus of the $C$ couplings involved in $B^0_{s}\to \tau^+\tau^-$ decay using Eq.~(\ref{eq:Cmutau}). We vary $\phi_{AN},\,\phi_{PP}^\tau$ from $0-\pi$ as there is no way we can constrain these phases.\\

As stated earlier that the experimental branching ratio may not be fully satisfied by SM only, there is some scope of NP. We explore all the possible scenarios. {\it At this point we would like to mention that in our following discussion whenever we talk about these couplings, we actually talk about their modulus.}\\

First in Case-I, we constrain the parameter space of the $C_{AA}^{SM}$ from the SM branching ratio of $B^0_{s}\to \mu^+\mu^-$\cite{sm-br} considering all NP terms as zero. From this constrain we can estimate the value of $A_\tau$ which satisfies the range from 19 to 21.5 for this case. Due to the absence of NP, the coefficient $B_\tau$ is zero here.\\

Next in Case-II, we include only $NP_1$ from all possible NP contributions with our SM contribution and compare the estimated branching ratio with the experimental branching ratio of $B^0_s\to \mu^+\mu^-$\cite{exp-b-mumu}. It gives bound on $C_{AA}$ which has been used to figure out the allowed range of $A_{\tau}$. Here $A_{\tau}$ ranges from $0.65-64.6$. Like that in Case-III, we choose only $NP_2$ coupling with SM. In this case $A_{\tau}$ ranges from $0.3-63.8$. These large ranges for $A_\tau$ in these two situations are due to the constructive and destructive interference between the SM and NP couplings. $\phi_{AN}$ and $\phi_{PP}^\tau$ are individually playing a key role to increase the range of $A_\tau$ from the sole presence of SM. In both of these situations, $B_\tau$ remains zero as we can have nonzero $B_\tau$ only if $NP_3$ is present. If we add only $NP_3$ with SM which is Case-IV, then the value of $A_\tau$ remains exactly same as Case-I. This is quite natural as $NP_3$ alone cannot affect the value of $A_\tau$. This situation is significantly different from Case-II and Case-III as in this case we have nonzero $B_\tau$ and it ranges between $0-11.4$.   \\

After this we consider situations where any two NP couplings are present with SM. In Case-V, we consider the presence of $SM + NP_1+NP_2$. Here $A_\tau$ ranges from $0.35-83.3$ and $B_\tau$ remains 0. In this case a very unnatural fine tuning happens between SM, $NP_1$ and $NP_2$. It gives unbounded parameter space for $NP_1$ and $NP_2$. It happens due to the presence of three interference terms between SM-$NP_1$, SM-$NP_2$ and $NP_1$-$NP_2$. In this situation, we take a logistic approach, where we neglect the values of the couplings for which fine tuning among SM, $NP_1$ and $NP_2$ is less then 20\%. This amount of fine tuning is sufficient to show the difference between the presence of NP from SM.  This choice of fine tuning is a very common practice in existing literature\cite{finetune}. In this way we avoid a very unnatural fine tuning between all the couplings. \\

Similarly we figure out the ranges for $A_\tau$ and $B_\tau$ for the cases of simultaneous presence of $SM + NP_1+NP_3$ and $SM + NP_2+NP_3$ which are mentioned in Case-VI and Case-VII respectively. All these values are listed in Table.~(\ref{tab:decaywidth}).  In Case-VIII we consider that all the NP couplings are present with SM. In that case $A_\tau$ ranges from $0.95-81.6$ and $B_\tau$ ranges from $0-22.8$. In Case-V to Case-VIII, we can notice that when both $NP_1$ and $NP_2$ are simultaneously present with SM, we get maximum range for $A_\tau$ almost$( 0 -81 )$. This is expected as for these two cases two phases $\phi_{AN}$ and $\phi_{PP}^\tau$ are simultaneously playing important role in the interference between various couplings. If only $\phi_{AN}$ or $\phi_{PP}^\tau$ is present then $A_\tau$ mostly remains between $0-65$ which is quite large compared to the case when only SM is present i.e. Case-I$(19-21.5)$. The role of the phase factors in the interference between different couplings is quite clear from all these situations. \\

\begin{figure*}[htbp]
\vspace{-10pt}
\centerline{\hspace{-3.3mm}
\rotatebox{-90}{\epsfxsize=6cm\epsfbox{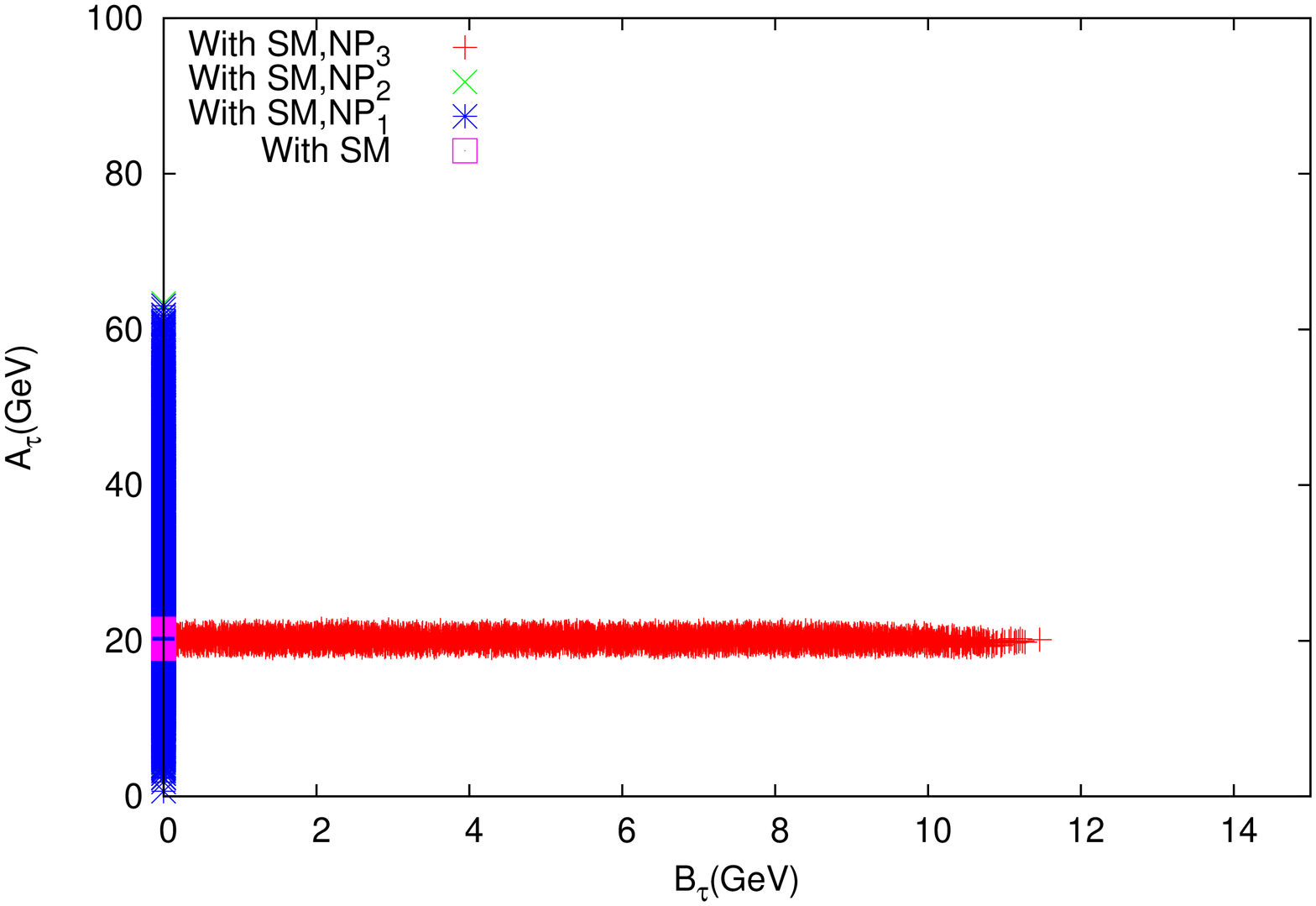}}
\hspace{0.3cm}\rotatebox{-90}{\epsfxsize=6cm\epsfbox{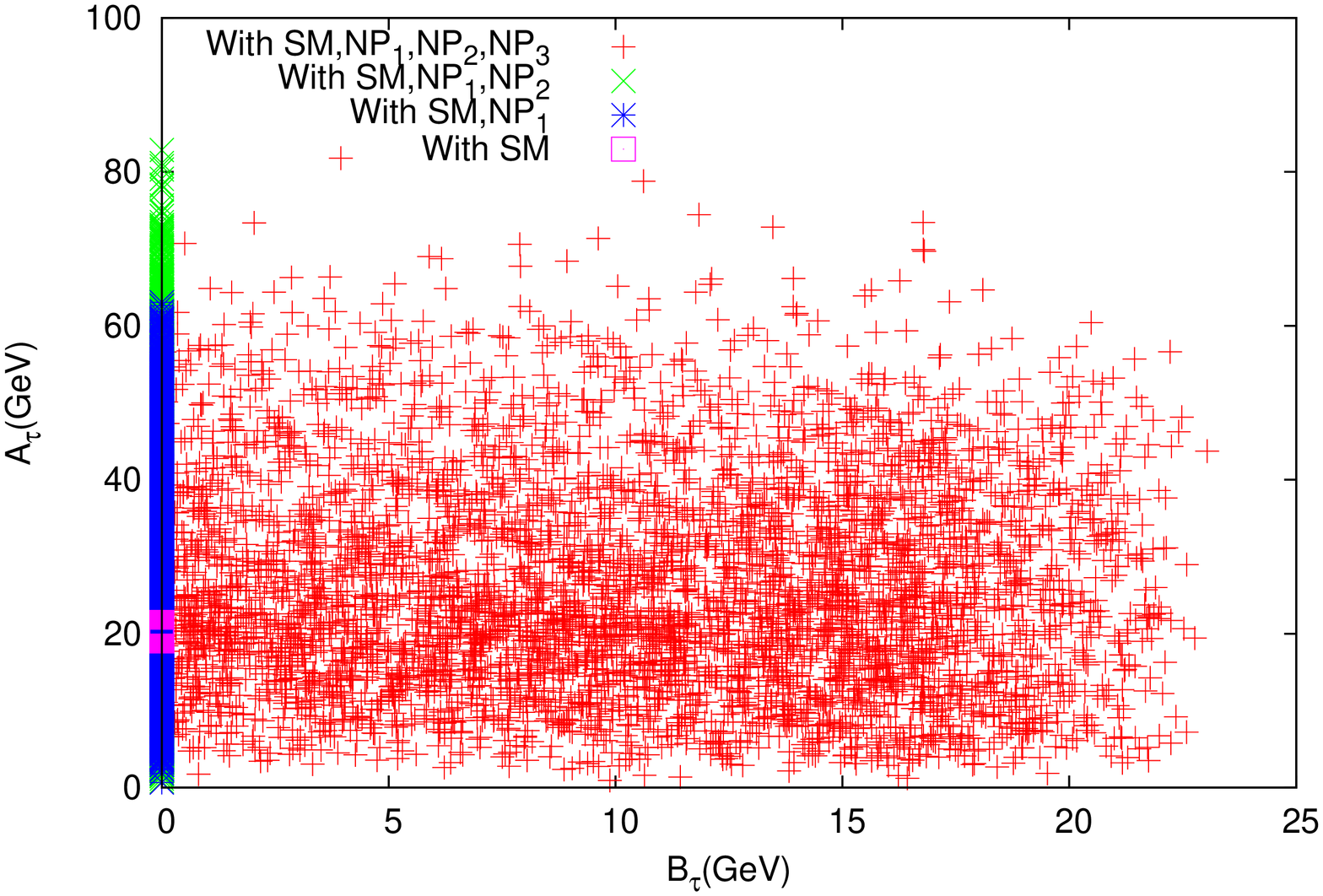}}}
\vspace*{3mm}
\centerline{\hspace{-0.5cm} (a) \hspace{7.5cm} (b)}
\hspace{3.3cm}
\caption{(a) Allowed ranges of $|A_\tau|$ and $|B_\tau|$ for Case-I, II, III and IV.
(b) Allowed ranges of $|A_\tau|$ and $|B_\tau|$ for Case-I, V, VI and VII.}
\label{fig:AB-13}
\end{figure*}

\begin{figure}
\rotatebox{-90}{\epsfxsize=6cm\epsfbox{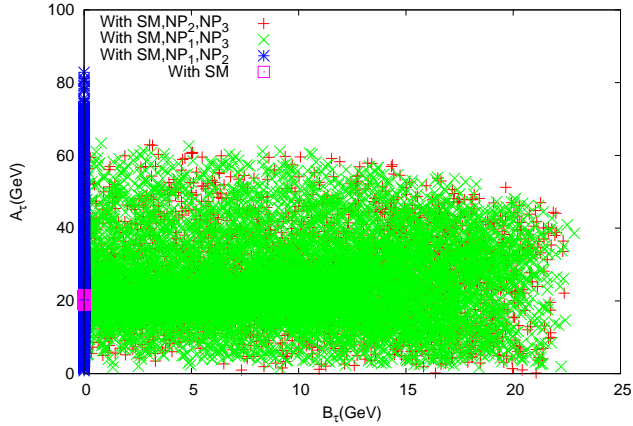}}
\caption{  Allowed ranges of $|A_\tau|$ and $|B_\tau|$ for Case-I, II, V and VIII.}
\label{fig:AB-2}
\end{figure}

All these cases and their outcomes are presented in Table.~(\ref{tab:decaywidth}), Fig.~(\ref{fig:AB-13}) and Fig.~(\ref{fig:AB-2}). At this point we want to mention that we have verified the allowed parameter space of $C_{AA}$, $C_{PP}$, $C_{PS}$, $A_\tau$ and $B_\tau $ using the time integrated SM value of branching ratio for $B_s^0 \to \mu^+\mu^-$ \cite{tism} too. We have not found any significant difference in the estimated values of $A_\tau$ and $B_\tau$. \\

Our main motivation is to isolate the NP from the SM so that it can be probed via experiment or at least establish a process of analysis such that the presence of any kind of NP can be identified prominently. It leads us to the obvious question that for what values of $A_{\tau}$ and $B_\tau$ we should continue our further analysis. Table.~(\ref{tab:decaywidth}) gives us the maximum allowed value for $A_{\tau}$ is 81.6 and $B_\tau$ is 22.8. The minimum is 0.2 and 0 respectively as mentioned in Table.~(\ref{tab:decaywidth}). In next section we discuss about the way to isolate NP from SM.

\begin{figure}
{\epsfxsize=8cm\epsfbox{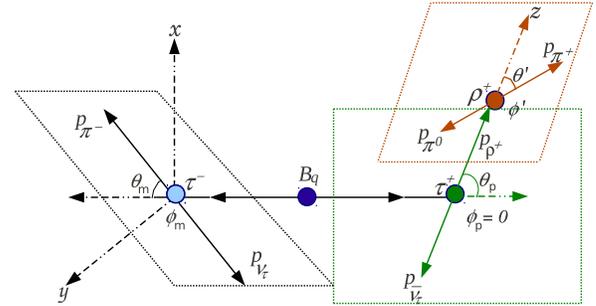}}
\caption{Four momentums and the rest frame of the intermediate particles of cascade decay. }
\label{fig:fig3b}
\end{figure} 

\section{ III. \boldmath The decay $B^0_s\to\tau^{+}\tau^{-}\to(\rho^+\bar{\nu}_\tau)_{\tau^+}(\pi^{-}\nu_{\tau})_{\tau^-}\to ((\pi^+\pi^0)_{\rho^+}+\bar{\nu}_\tau)_{\tau^+}(\pi^{-}\nu_{\tau})_{\tau^-}$}
\label{sec:rho-pi}

It is now clear that though experimental branching ratio of $B^0_{s}\to \mu^+\mu^-$ is within the same regime of SM, the possibility of the presence of NP is not ruled out. However, the main challenge is how to experimentally observe the presence of NP. For this purpose we are exploring a good old technique of angular analysis\cite{angular-analysis}. For such angular analysis we will choose $B^0_{s}\to \tau^+\tau^-$ not the $B^0_{s}\to \mu^+\mu^-$ as further decays of muons is not suitable for doing angular analysis to isolate the NP from SM. The decay of $B^0_{s}\to \tau^+\tau^-$ is also governed by the same effective Hamiltonian involved in $B^0_{s}\to \mu^+\mu^-$ decay except for the fact that $\mu$ mass is replaced by $\tau$. Then we consider the further decays of $\tau$'s. It makes $B^0_s\to\tau^{+}\tau^{-}\to(\rho^+\bar{\nu}_\tau)_{\tau^+}(\pi^{-}\nu_{\tau})_{\tau^-}\to ((\pi^+\pi^0_{\rho^+}+\bar{\nu}_\tau)_{\tau^+}(\pi^{-}\nu_{\tau})_{\tau^-}$ as an interesting choice to perform angular analysis. In this section our primary motivation is to present a technique to  distinguish experimentally the various NP effects from the SM contribution through angular analysis of this cascade decay channel. \\

We divide this section into two parts. For the cascade decay, we provide the information about different reference frames, momentums of various particles involved, assumptions of ignoring negligible masses, constraints from energy-momentum conservation relations etc. in the first part. Second part involves detail analysis. Where we show that how NP can be isolated from SM via this analysis. However, one very essential and remarkable thing which we establish there is that if any kind of NP is present then that situation is significantly different from the sole presence of SM in our angular analysis, which can be tested experimentally at LHCb, using sophisticated technique.  
    
\subsection{ III. A. Relevant formulae for the cascade decay of $B_s^0$ }
      
We begin by considering the decay $B^0_s(P)\to\tau^+(p_+)\tau^-(p_-)$. And then consider, the $\tau^+$ and $\tau^-$ to decay further with $\tau^+(p_+)\to\rho^+(q)\nu_\tau(q_{\bar{\nu}})$ and
$\tau^-(p_-)\to \pi^-(k) \nu_\tau(k_\nu)$. The $\rho^+$ produced is considered to decay into $\pi^+$ and $\pi^0$, with four-momentum $q_1$ and $q_2$ respectively. We first define the kinematics of the process by expressing the four momentum vectors of all the particles in the decay process. We describe the decay in the $B_s^0$ rest frame, where, the $\tau^+$ and $\tau^-$ decay back to back. The $z$ axis is defined to be along the direction of the $\rho^+$ as shown in Fig.~\ref{fig:fig3b}.

The momenta and the angles involved in the decay are related in the $B^0_s$ rest frame as follows:
\bea
\label{eq:Bresteq}
P^{\mu}(B_s^0)&=&\{~M, 0, 0, 0~\},\n \\
p_+^{\mu}(\tau^+)&=&\frac{M}{2}\{1,y \sin
    \theta_{\tau}, 0, y \cos \theta_{\tau}~\}, \n \\ 
p_-^{\mu}(\tau^-)&=&\frac{M}{2}(1, -y \sin
    \theta_{\tau}, 0, -y \cos \theta_{\tau}~\}, \n \\ 
q^{\mu}(\rho^+) &=&\{~\sqrt{q^2+m_{\rho}^2}, 0, 0, q~\},\n \\ 
k^{\mu}(\pi^-)&=&\{~k, k \sin \theta \cos \phi, k
    \sin \theta \sin\phi, k \cos \theta ~\}, \n \\ 
q_1^{\mu}(\pi^+)&=&\{~q_1, q_1 \sin \theta_1 \cos
    \phi_1, q_1 \sin \theta_1 \sin \phi_1, q_1 \cos \theta_1 ~\}, \n 
\eea
where we have used $P^{\mu} = p_+^{\mu} + p_-^{\mu}$ to obtain 
\begin{equation}
  \label{eq:y}
  y=\Big(1-\frac{4\,{m_{\tau}}^2}{M^2}\Big)^{\frac{1}{2}}.
\end{equation}
The remaining three energy-momentum conservation relations, $p_+^{\mu} = q^{\mu} + q_{\sss\bar{\nu}}^{\mu}$, $p_-^{\mu} = k^{\mu} + k_{\sss\nu}^{\mu}$, and $q^{\mu} = q_1^{\mu} + q_2^{\mu}$, give us the freedom to eliminate $q_{\sss\bar{\nu}}^{\mu}, k_{\sss\nu}^{\mu}, q_2^{\mu}$ in terms of the rest of the momentums. The on-shell conditions for the initial $B^0_s$ meson, the final state pions and the neutrinos are $P^2=M^2$, $q_1^2=q_2^2=k^2=m_\pi^2$ and $q_{\bar{\nu}}^2=k_{\nu}^2=0$ respectively. We also require the intermediate $\tau^\pm$ leptons and $\rho$ meson to be on shell, which is imposed by using $p_+^2=p_-^2=m_\tau^2$ and $q^2=m_\rho^2$ respectively. Energy-momentum conservation gives $(p_+^{\mu}- q^{\mu})^2=0=m_\tau^2+m_\rho^2-M\sqrt{q^2+m_\rho^2}+M q y \cos\theta$, resulting in a relation for $\theta_\tau$:
\begin{equation}
  \label{eq:theta_tau}
\cos \theta_{\tau}= \frac{1}{q y}\left[~ \sqrt{q^2 + m_{\rho}^2} - 
\frac{1}{M}(m_{\tau}^2 + m_{\rho}^2) ~\right]~.  
\end{equation}

The $\pi^-$ and $\nu_\tau$ momentum can be easily written in $\tau^-$ rest frame since $\pi^-$ and $\nu_{\tau}$ are back to back. Neglecting the masses of pions and neutrinos, their respective four momentums are as follows : \\
\bea
\label{eq:tau-m-rest}
p_{\pi^-}^{\mu} &=& \frac{m_{\tau}}{2}\{1,\sin\theta_{\!\sss \rm m}\cos\phi_{\!\sss \rm m},\sin\theta_{\!\sss \rm m}\sin\phi_{\!\sss \rm m},\cos\theta_{\!\sss \rm m} \},\\
p_{\sss\nu}^{\mu} &=& \frac{m_{\tau}}{2}\{1,-\sin\theta_{\!\sss \rm m}\cos\phi_{\!\sss \rm m},-\sin\theta_{\!\sss \rm m}\sin\phi_{\!\sss \rm m},-\cos\theta_{\!\sss \rm m} \}. \n
\eea
We note that $\theta_{\!\sss \rm m}$ and $\phi_{\!\sss \rm m}$ are angles described in the $\tau^-$ rest frame. Similarly, in $\tau^+$ rest frame, $\rho^+$ and $\bar{\nu}$ are back to back and their respective four momentums are as given in terms of $\tau^+$ rest frame angles $\theta_{\!\sss \rm p}$ and $\phi_{\!\sss \rm p}$ as follows:\\
\bea
\label{eq:tau-p-rest1}
\begin{split}
p_{\rho^+}^{\mu}= \{\frac{m_{\tau}^2 + m_{\rho}^2}{2 m_{\tau}},\frac{m_{\tau}^2 - m_{\rho}^2}{2 m_{\tau}}\sin\theta_{\!\sss
  \rm p}\cos\phi_{\!\sss \rm p},\\
 \frac{m_{\tau}^2 - m_{\rho}^2}{2 m_{\tau}}\sin\theta_{\!\sss \rm
  p}\sin\phi_{\!\sss \rm p},\frac{m_{\tau}^2 - m_{\rho}^2}{2 m_{\tau}}\cos\theta_{\!\sss \rm p} ~\}
\end{split}
\eea
\bea
\label{eq:tau-p-rest2}
p_{\sss\bar{\nu}}^{\mu} =
\frac{m_{\tau}^2 - m_{\rho}^2}{2 m_{\tau}}\{1,-\sin\theta_{\!\sss \rm p}\cos\phi_{\!\sss \rm
  p},-\sin\theta_{\!\sss \rm p}\sin\phi_{\!\sss \rm
  p},-\cos\theta_{\!\sss \rm p} ~\}. \n 
\eea

With the same logic the $\pi^+$ and $\pi^0$ are back to back in the $\rho^+$ rest frame. Considering $\rho^+$ along z axis and neglecting masses of pions and neutrinos, their respective four momentums are described in terms of the rest frame angles $\theta^{\prime}$ and $\phi^{\prime}$ as follows:\\
\bea
\label{eq:rho-p-rest}
p_{\pi^+}^{\mu} &=& \frac{m_{\rho}}{2}\{1,\sin\theta^{\prime}\cos\phi^{\prime},\sin\theta^{\prime}\sin\phi^{\prime},\cos\theta^{\prime}\},\\
p_{\pi^0}^{\mu} &=& \frac{m_{\rho}}{2}\{1,-\sin\theta^{\prime}\cos\phi^{\prime},-\sin\theta^{\prime}\sin\phi^{\prime},-\cos\theta^{\prime} ~\}.\n
\eea

The angles of decay products defined in the $B^0_s$ rest frame can be expressed in terms of the angles defined in the respective rest frames of $\tau^+$ ($\theta_{\!\sss\rm p}$, $\phi_{\!\sss \rm p}$), $\tau^-$ ($\theta_{\!\sss\rm m}$, $\phi_{\!\sss \rm m}$) and $\rho^+$ ($\theta^{\prime}$, $\phi^{\prime}$). We always have the freedom to choose either $\phi_{\!\sss \rm p}$ or $\phi_{\!\sss \rm m}$ as zero as the relative angle between any two azimuthal angles is the only relevant quantity. Without any loss of generality we choose $\phi_{\!\sss \rm p} = 0$. The relations between the other two azimuthal angles are simplified to $\phi = \phi_{\sss \rm m}$ and $\phi_1 = \phi^{\prime}$ with this choice. The remaining relations are,
\begin{widetext}
\bea
\label{eq:lab-rest-relation}
\cos \theta &=& -\frac{M}{4 k}\left[~(y-\cos \theta_{\!\sss \rm m})\cos\theta_{\tau} + \sqrt{1-y^2}\sin\theta_{\tau}\sin\theta_{\!\sss \rm m}\cos\phi_{\!\sss \rm m} ~\right] ,\n \\
\cos \theta_1 &=& \frac{1}{2 q_1} \left[~ q + \sqrt{q^2 + m_{\rho}^2}\cos\theta^{\prime} ~\right],\n \\
q &=& \frac{M}{4 m_{\tau}^2}\left[~ \{ (m_{\tau}^2+m_{\rho}^2) +y(m_{\tau}^2-m_{\rho}^2)\cos\theta_{\!\sss \rm p} \}^2 - \frac{16 m_{\tau}^4 m_{\rho}^2}{M^2} ~\right]^{\frac 1 2}, \n \\
k &=& \frac{M}{4}\left[ ~ 1 - y\cos\theta_{\!\sss \rm m} ~\right],  \n \\
q_1 &=& \frac 1 2 \left[ ~ \sqrt{q^2 + m_{\rho}^2} + q \cos\theta^{\prime} ~\right]. \n 
\eea
The partial decay width for $ B_s^0 \rightarrow \tau^+ \tau^- \rightarrow \rho^+ \bar{\nu_{\tau}} \pi^{-} \nu_{\tau} \rightarrow \pi^+ \pi^0 \bar{\nu_{\tau}} \pi^{-} \nu_{\tau} $ can be written as
\bea
\label{eq:d-phase-space}
d\Gamma &=& \frac{3y ( \alpha \ f_{B_q} \ G_F)^2 \ |V_{tq}^*|^2  |V_{tb}|^2 }{64 \ \pi^5 \ M \ m_{\tau}^8 } \mathcal{BR}\left(~\tau^+ \rightarrow \rho^+ \bar{\nu}) \mathcal{BR}(\rho^+\rightarrow \pi^+ \pi^0 ~\right) \mathcal{BR}\left(~\tau^- \rightarrow \pi^- \nu_{\tau} ~\right) \n \\
&& |{\cal{M}}|^2 \ d(\cos\theta_{\!\sss \rm p}) \ d(\cos\theta_{\!\sss \rm m}) \ d(\cos\theta^{\prime}) \ d(\phi_{\!\sss \rm m}) \ d(\phi^{\prime}).
\eea
\end{widetext}

At this point we would like to mention that $B_s^0$ rest frame can be chosen as lab frame and angles in this frame are measurable quantities. Matrix element for this cascade decay is expressed in terms of these lab angles in Table.(III) - (VI), whereas from Eq.~(\ref{eq:d-phase-space}) it is clear that $\tau^+$, $\tau^-$, $\rho^+$ rest frame angles are necessary to numerically estimate the total decay width $\Gamma$ for this cascade decay, though these angles cannot be measured. For this purpose we have provided the relations between lab and individual rest frame angles in this section. The final matrix element square for the cascade decay of $B_s^0$ is given in Appendix.

\subsection{ III. B. Analysis}

This cascade decay involves total five independent angles, 3 polar which can be measured without any ambiguity and two azimuthal angles which can be measured with two fold ambiguity. With these angle information, we prescribe five observables as $\frac{d\Gamma}{dcos\theta_\tau}, \frac{d\Gamma}{dcos\theta}, \frac{d\Gamma}{dcos\theta_1}, \frac{d\Gamma}{d\phi_1}, \frac{d\Gamma}{d\phi}$. These observables are functions of $A_\tau$ and $B_\tau$, hence, sensitive to various NP. Using the constraints obtained for $A_\tau$ and $B_\tau$ from Table.~(\ref{tab:decaywidth}), we numerically estimate the sizes of these partial decay widths with the help of Vegas\cite{vegas}. We have estimated that at 14TeV, with 50$fb^{-1}$ total integrated luminosity, LHCb can generate $10^4 - 10^7$ $B^0_{s}\to \tau^+\tau^-$ events for our allowed range of $A_\tau$ and $B_\tau$. So there will be sufficient number of events available to verify our analysis.\\

$A_\tau$ involves both SM and NP terms whereas $B_\tau$ involves purely NP term. We divide our analysis into two separate parts. In the first part, our main motivation is to prescribe a way to detect the the presence of NP. After that we try to show that if NP is present then ``how do we determine it's Lorentz structure?". In the second part of our analysis, we explore the region where these partial decay widths are sensitive to $B_\tau$ for a fixed value of $A_\tau$. This sensitivity gives us a hope to isolate a pure NP term from SM experimentally. 

\subsubsection{\bf Definite indication of New Physics }

\begin{figure}
{\epsfxsize=8cm\epsfbox{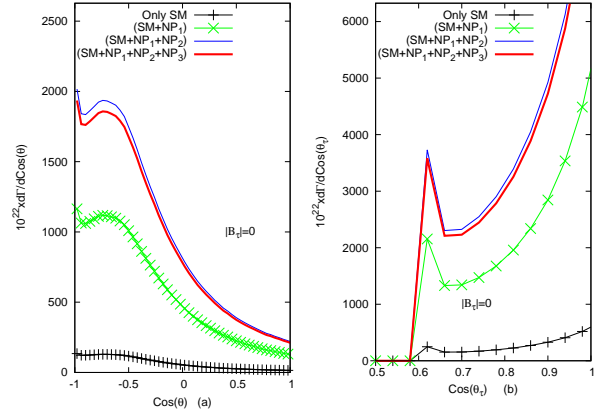}}
\caption{(a) Variation of observable $\frac{d\Gamma}{dCos\theta}$ as a function of $Cos\theta$ for SM, SM+$NP_1$, SM+$NP_1$+$NP_2$ and SM+$NP_1$+$NP_2$+$NP_3$ for largest allowed $|A_\tau|$.
(b) Variation of observable $\frac{d\Gamma}{dCos\theta_\tau}$ as a function of $Cos\theta_\tau$ for SM, SM+$NP_1$, SM+$NP_1$+$NP_2$ and SM+$NP_1$+$NP_2$+$NP_3$ for largest allowed $|A_\tau|$. 
For all these diagrams we have chosen $|B_\tau|=0$.}
\label{fig:fig0}
\end{figure} 

\begin{figure}
{\epsfxsize=8cm\epsfbox{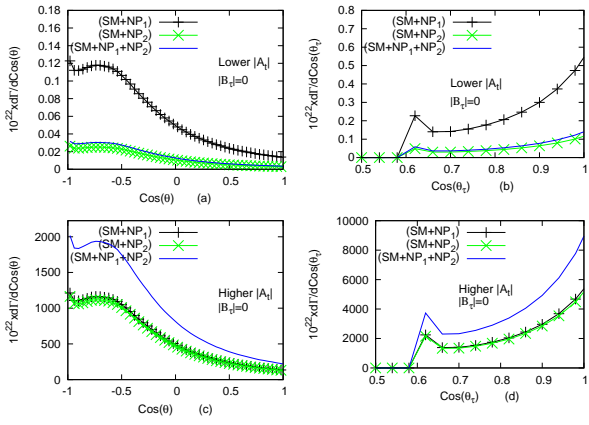}}
\caption{(a) Variation of observable $\frac{d\Gamma}{dCos\theta}$ as a function of $Cos\theta$ for different values of $|A_\tau|$ (smallest values).
(b) Variation of observable $\frac{d\Gamma}{dCos\theta_\tau}$ as a function of $Cos\theta_\tau$ for different values of $|A_\tau|$ (smallest values). 
(c) Variation of observable $\frac{d\Gamma}{dCos\theta}$ as a function of $Cos\theta$ for different values of $|A_\tau|$ (largest values).
(d) Variation of observable $\frac{d\Gamma}{dCos\theta_\tau}$ as a function of $Cos\theta_\tau$ for different values of $|A_\tau|$ (largest values).
For all these diagrams we have chosen $|B_\tau|=0$.}
\label{fig:fig2}
\end{figure} 

\begin{figure}
{\epsfxsize=8cm\epsfbox{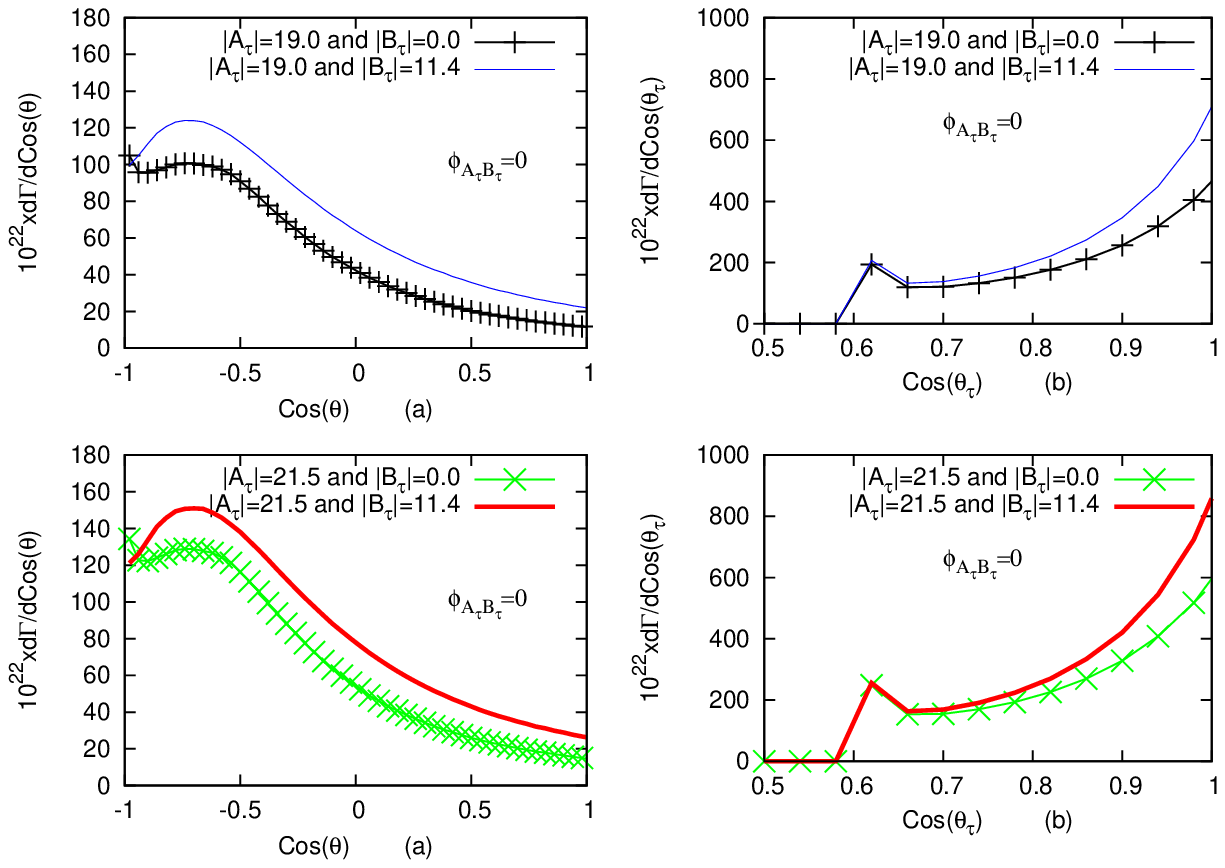}}
\caption{(a) Variation of observable $\frac{d\Gamma}{dCos\theta}$ as a function of $Cos\theta$ for smallest value of $|A_\tau|$ and smallest and largest values of $|B_\tau|$ for Case-IV.
 (b) Variation of observable $\frac{d\Gamma}{dCos\theta_\tau}$ as a function of $Cos\theta_\tau$ for smallest value of $|A_\tau|$ and smallest and largest values of $|B_\tau|$ for Case-IV.
(c) Variation of observable $\frac{d\Gamma}{dCos\theta}$ as a function of $Cos\theta$ for largest value of $|A_\tau|$ and smallest and largest values of $|B_\tau|$ for Case-IV.
 (d) Variation of observable $\frac{d\Gamma}{dCos\theta_\tau}$ as a function of $Cos\theta_\tau$ for largest value of $|A_\tau|$ and smallest and largest values of $|B_\tau|$ for Case-IV.
For both of these diagrams we have chosen the relative phase $\phi_{A_\tau\,B_\tau}$ between $A_\tau$ and $B_\tau$ as zero.}
\label{fig:fig3}
\end{figure}

\begin{figure}
{\epsfxsize=8cm\epsfbox{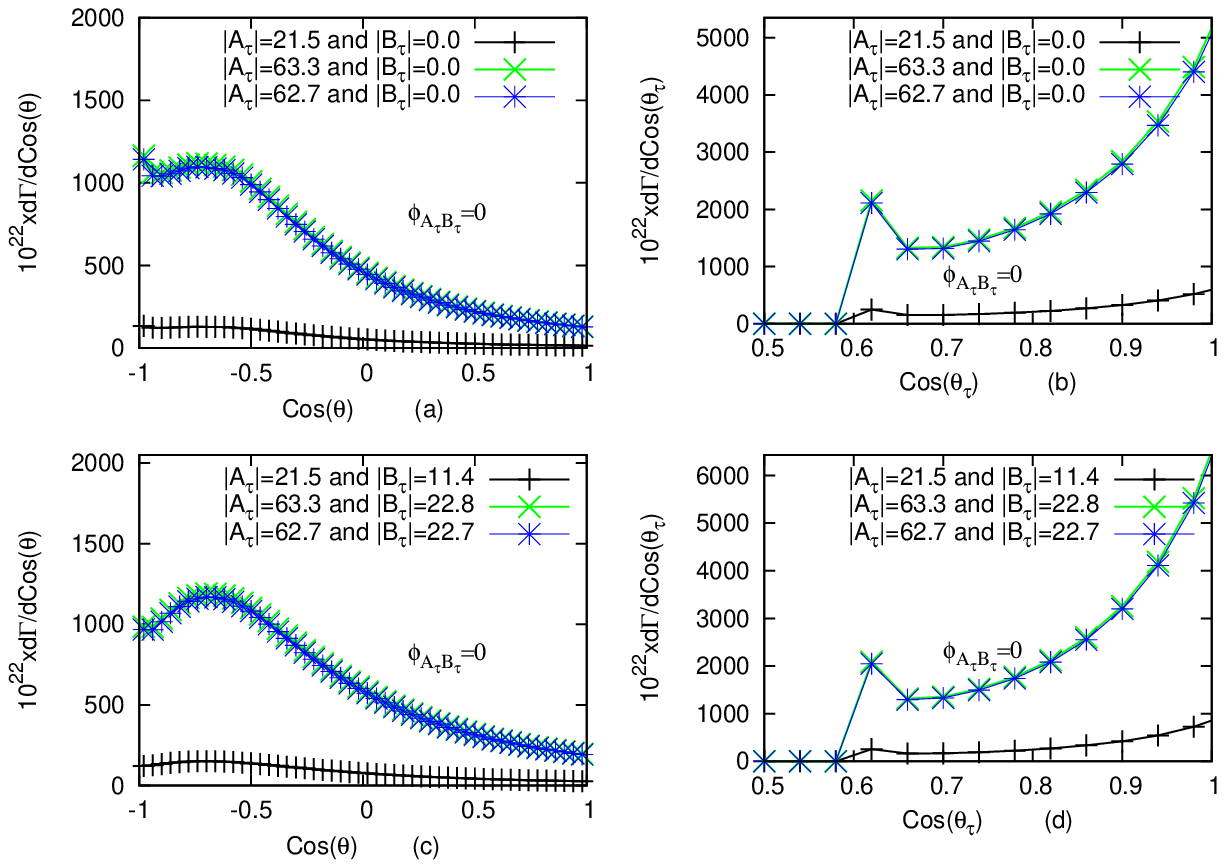}}
\caption{(a) Variation of observable $\frac{d\Gamma}{dCos\theta}$ as a function of $Cos\theta$ for largest value of $|A_\tau|$ and for smallest value of $|B_\tau|$ for Case-IV, VI and VII.
(b) Variation of observable $\frac{d\Gamma}{dCos\theta_\tau}$ as a function of $Cos\theta_\tau$ for largest value of $|A_\tau|$ and for smallest values of $|B_\tau|$ for Case-IV, VI and VII.
(c) Variation of observable $\frac{d\Gamma}{dCos\theta}$ as a function of $Cos\theta$ for largest values of $|A_\tau|$ and $|B_\tau|$ for Case-IV, VI and VII.
(d) Variation of observable $\frac{d\Gamma}{dCos\theta_\tau}$ as a function of $Cos\theta_\tau$ for largest values of $|A_\tau|$ and $|B_\tau|$ for Case-IV, VI and VII.
For all these diagrams we have chosen the relative phase $\phi_{A_\tau\,B_\tau}$ between $A_\tau$ and $B_\tau$ as zero. }
\label{fig:fig5}
\end{figure} 

\begin{figure}
{\epsfxsize=8cm\epsfbox{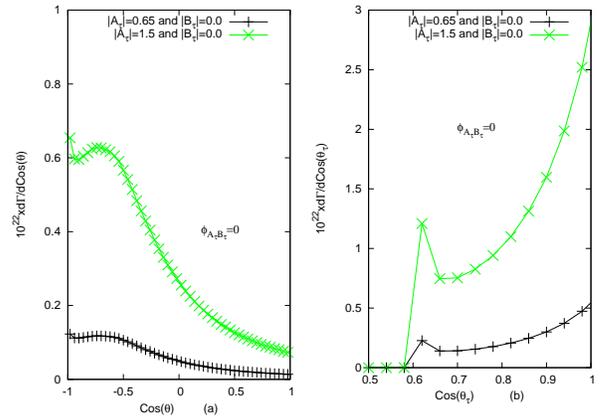}}
\caption{(a) Variation of observable $\frac{d\Gamma}{dCos\theta}$ as a function of $Cos\theta$ for smallest values of $|A_\tau|$ and $|B_\tau|$ for Case-II and VI.
(b) Variation of observable $\frac{d\Gamma}{dCos\theta_\tau}$ as a function of $Cos\theta_\tau$ for smallest values of $|A_\tau|$ and $|B_\tau|$ for Case-II and VI.}
\label{fig:fig6}
\end{figure}

In this section our main aim is to show whether we can clearly establish the presence of NP or not. For this purpose we take two approaches.\\
\begin{itemize}
\item[1.] The first approach is to estimate the decay width $\Gamma$ of the cascade decay $B^0_{s}\to \tau^+\tau^- \to (\rho^+\bar{\nu}_\tau)_{\tau^+}(\pi^{-}\nu_{\tau})_{\tau^-} \to ((\pi^+ \pi^0)_{\rho^+}\bar{\nu}_\tau)_{\tau^+}(\pi^{-}\nu_{\tau})_{\tau^-}$ and to check whether this $\Gamma$ can tell us anything about NP or not. 
\item[2.] The second approach is to do an angular analysis of the same cascade decay with five partial decay widths $\frac{d\Gamma}{dcos\theta_\tau}, \frac{d\Gamma}{dcos\theta}, \frac{d\Gamma}{dcos\theta_1}, \frac{d\Gamma}{d\phi_1}, \frac{d\Gamma}{d\phi}$ which are functions of $\theta_\tau$, $\theta$, $\theta_1$, $\phi_1$ and $\phi$ respectively. For this second approach we investigate whether these partial decay widths can provide us more information about the presence of NP or not.
\end{itemize}

To figure out $\Gamma$ for various cases mentioned in Table.~(\ref{tab:decaywidth}), first we consider $B_\tau$ as zero and estimate the ranges of $\Gamma$ for allowed ranges of $A_\tau$. These constrained ranges are mentioned in the fourth column of Table.~(\ref{tab:decaywidth}). Next we explore the possibility of including the allowed range of $A_\tau$ in presence of allowed maximum $B_\tau$ to figure out $\Gamma$ and list them in the last column of Table.~(\ref{tab:decaywidth}). From this Table.~(\ref{tab:decaywidth}), it is clear that if any NP is present, $\Gamma$ differs by at least one order from Case-I i.e. from the situation where only SM is present. It means that the estimation of the decay width of the cascade decay is sufficient to detect the presence of NP. Though the main problem with this approach is that it is beyond the present experimental limit to observe these decay width as these are extremely small. On top of that estimation of $\Gamma$ cannot distinguish the nature of the NP. \\

In our second approach, we perform angular analysis. We estimate the partial decay width $\frac{d\Gamma}{dcos\theta}, \frac{d\Gamma}{dcos\theta_\tau}$ as a function of $cos\theta$ and $cos\theta_\tau$ respectively for SM, SM+$NP_1$, SM+$NP_1$+$NP_2$ and SM+$NP_1$+$NP_2$+$NP_3$ for largest allowed $A_\tau$ and $B_\tau$ as zero using Vegas and present it in Fig.~\ref{fig:fig0}(a)-(b). From this figure it is extremely clear that presence of any kind of NP can be clearly established from the case where only SM is present from the estimation of these partial decay widths. We do not present similar figure for the smallest ranges of $A_\tau$ though we have explored that situation too. In that case the plot for $\frac{d\Gamma}{dcos\theta}, \frac{d\Gamma}{dcos\theta_\tau}$ in the presence of NP differ by almost 2 order from the sole presence of SM, which makes our analysis even stronger. We have performed similar analysis for the rest of the partial decay widths $\frac{d\Gamma}{dcos\theta_1}, \frac{d\Gamma}{d\phi_1}, \frac{d\Gamma}{d\phi}$ and have come to similar conclusion, though we do not present explicit plots for these observables in this paper.  {\it At this point we would like to  mention that without any loss of generality we have scaled $\frac{d\Gamma}{dcos\theta}$ and $\frac{d\Gamma}{dcos\theta_\tau}$ by a factor of $10^{22}$ in Fig.~\ref{fig:fig0} - Fig.~\ref{fig:fig8} due to the smallness of these variables. }\\

The biggest advantage of this angular analysis is that we can measure the five angles $\theta$, $\theta_\tau$, $\theta_1$, $\phi$ and $\phi_1$. The measurement of these angles can be used to estimate the partial decay widths $\frac{d\Gamma}{dcos\theta_\tau}, \frac{d\Gamma}{dcos\theta}, \frac{d\Gamma}{dcos\theta_1}, \frac{d\Gamma}{d\phi_1}, \frac{d\Gamma}{d\phi}$ which are functions of $\theta_\tau$, $\theta$, $\theta_1$, $\phi_1$ and $\phi$ respectively. These experimental estimation can be compared with our theoretical estimation and presence of NP can be established without much ambiguity.\\

In future if experimentally it is established that NP is present with SM using our analysis, the next immediate question which will rise is ``what is the nature of this NP?" We present an analysis which can answer this question too. First we assume that $NP_3$ is absent. Then the remaining NP are $NP_1$ and $NP_2$. They can be present individually or simultaneously with SM. Case-II, III and V explore the possibilities of these situations. Now our job is to suggest some analysis through which we can differentiate these situations from each other. We can see from Table.~(\ref{tab:decaywidth}) that the allowed lowest value of $A_\tau$ in the sole presence of SM is 19 which shifts to 0.65, 0.3 and 0.35 for Case-II, III and V respectively. Similarly the allowed highest value of $A_\tau$ in the sole presence of SM is 21.5 which shifts to 64.6, 63.8 and 83.3 for Case-II, III and V respectively. The reason for it is already mentioned in Section-II. In all these three cases $B_\tau=0$. We plot $\frac{d\Gamma}{dcos\theta}$ as a function of $\cos\theta$ and for $A_\tau=0.65,\,0.3\,$ and $0.35$ and $B_\tau=0$ in Fig.~\ref{fig:fig2}(a). It shows a clear difference in the variable $\frac{d\Gamma}{dcos\theta}$ for Case-II and Case-III though Case-V looks very similar to Case-III as in these two cases the lowest values of $A_\tau$ are quite similar. Similar plot is presented in Fig.~\ref{fig:fig2}(b) for $\frac{d\Gamma}{dcos\theta_\tau}$ as a function of $cos\theta_\tau$. Here also we can distinguish Case-II from Case-III though Case-III and Case-V cannot be distinguished from each other. But this problem can be solved if we concentrate on Fig.~\ref{fig:fig2}(c) which has been plotted for the allowed highest values of $A_\tau=64.6, \, 63.8,\,83.3$ and $B_\tau=0$ for $\frac{d\Gamma}{dcos\theta}$. In this case though Case-II and Case-III are looking very similar but Case-V can be clearly distinguishable from Case-II and Case-III. Same conclusion goes for Fig.~\ref{fig:fig2}(d) which is a plot for $\frac{d\Gamma}{dcos\theta_\tau}$ for the allowed highest values of $A_\tau=64.6, \, 63.8,\,83.3$ and $B_\tau=0$. The analysis of these four plots gives us a clear picture of the way to distinguish the individual presence of $C_{AA}^{NP}$, $C_{PP}$ and simultaneous presence of both $C_{AA}^{NP}$ and $C_{PP}$ with SM. \\

Next we need to explain how to figure out the presence of $C_{PS}$ i.e. $NP_3$ with SM. From Table.~(\ref{tab:decaywidth}), we notice that though the ranges of $A_\tau$ are same for Case-I and Case-IV, the main difference comes in the allowed non-zero values of $B_\tau$ in Case-IV. This situation is explored in Fig.~\ref{fig:fig3}. Fig.~\ref{fig:fig3}(a),(c) represents the plot of $\frac{d\Gamma}{dcos\theta}$ as a function of $cos\theta$ and Fig.~\ref{fig:fig3}(b),(d) represents the plot of $\frac{d\Gamma}{dcos\theta_\tau}$ as a function of $cos\theta_\tau$. The detail values of $A_\tau$, $B_\tau$ and their relative phase $\phi_{A\tau\,B_\tau}$ are mentioned in the plots. From these four figures we can see that it requires precision measurement to determine the presence of $C_{PS}$ with SM. $NP_3$ is present in Case-VI, VII and VIII too. Now we have to check how to differentiate these cases from each other through angular analysis. Similar type of analysis is presented in Fig.~\ref{fig:fig5}(a)-(d) where Case-IV can be clearly distinguished from Case-VI and VII. How to separate Case-II from Case-VI, that is shown in Fig.~\ref{fig:fig6}(a)-(b). Like this, we can have many more possibilties to show that each of the cases mentioned in Table.~(\ref{tab:decaywidth}) can be distinguished from each other through angular analysis. We are not presenting all these figures. But it is now clear that using our angular analysis we not only can suggest a clean way to check the presence of NP, our analysis also has the potential to predict the Lorentz nature of the present NP.   \\


\subsubsection{\bf Isolate the New Physics from Standard Model}

\begin{figure}
{\epsfxsize=8cm\epsfbox{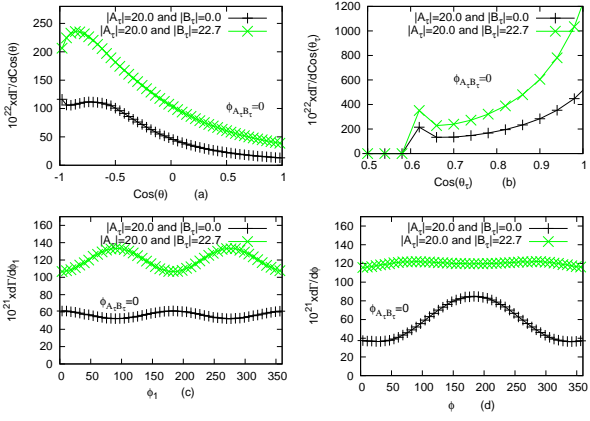}}
\caption{(a) Variation of observable $\frac{d\Gamma}{dCos\theta}$ as a function of $Cos\theta$ for $|A_\tau|=20$ and smallest and largest values of $|B_\tau|$ for Case-VI, VII and VIII.
(b) Variation of observable $\frac{d\Gamma}{dCos\theta_\tau}$ as a function of $Cos\theta_\tau$ for $|A_\tau|=20$ and smallest and largest values of $|B_\tau|$ for Case-VI, VII and VIII.
(c) Variation of observable $\frac{d\Gamma}{d\phi_1}$ as a function of $\phi_1$ for $|A_\tau|=20$ and smallest and largest values of $|B_\tau|$ for Case-VI, VII and VIII.
(d) Variation of observable $\frac{d\Gamma}{d\phi}$ as a function of $\phi$ for $|A_\tau|=20$ and smallest and largest values of $|B_\tau|$ for Case-VI, VII and VIII.
For all these diagrams we have chosen the relative phase $\phi_{A_\tau\,B_\tau}$ between $A_\tau$ and $B_\tau$ as zero.}
\label{fig:fig7}
\end{figure}

\begin{figure}
{\epsfxsize=8cm\epsfbox{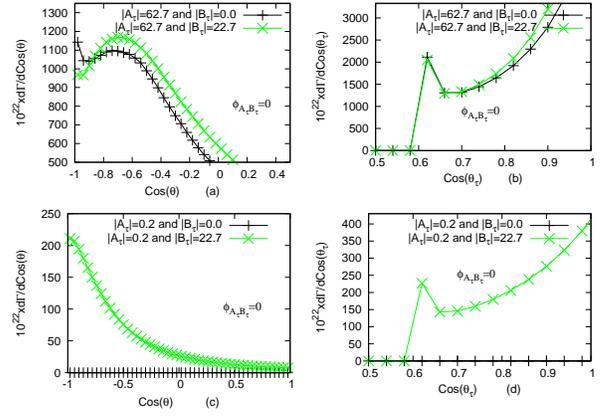}}
\caption{(a) Variation of observable $\frac{d\Gamma}{dCos\theta}$ as a function of $Cos\theta$ for largest value of $|A_\tau|$ and for smallest and largest values of $|B_\tau|$ for Case-VII.
(b) Variation of observable $\frac{d\Gamma}{dCos\theta_\tau}$ as a function of $Cos\theta_\tau$ for largest value of $|A_\tau|$ and for smallest and largest values of $|B_\tau|$ for Case-VII. 
(c) Variation of observable $\frac{d\Gamma}{dCos\theta}$ as a function of $Cos\theta$ for smallest value of $|A_\tau|$ and for smallest and largest values of $|B_\tau|$ for Case-VII.
(d) Variation of observable $\frac{d\Gamma}{dCos\theta_\tau}$ as a function of $Cos\theta_\tau$ for smallest value of $|A_\tau|$ and for smallest and largest values of $|B_\tau|$ for Case-VII.
For all these diagrams we have chosen the relative phase $\phi_{A_\tau\,B_\tau}$ between $A_\tau$ and $B_\tau$ as zero.}
\label{fig:fig4}
\end{figure}  

\begin{figure}
{\epsfxsize=8cm\epsfbox{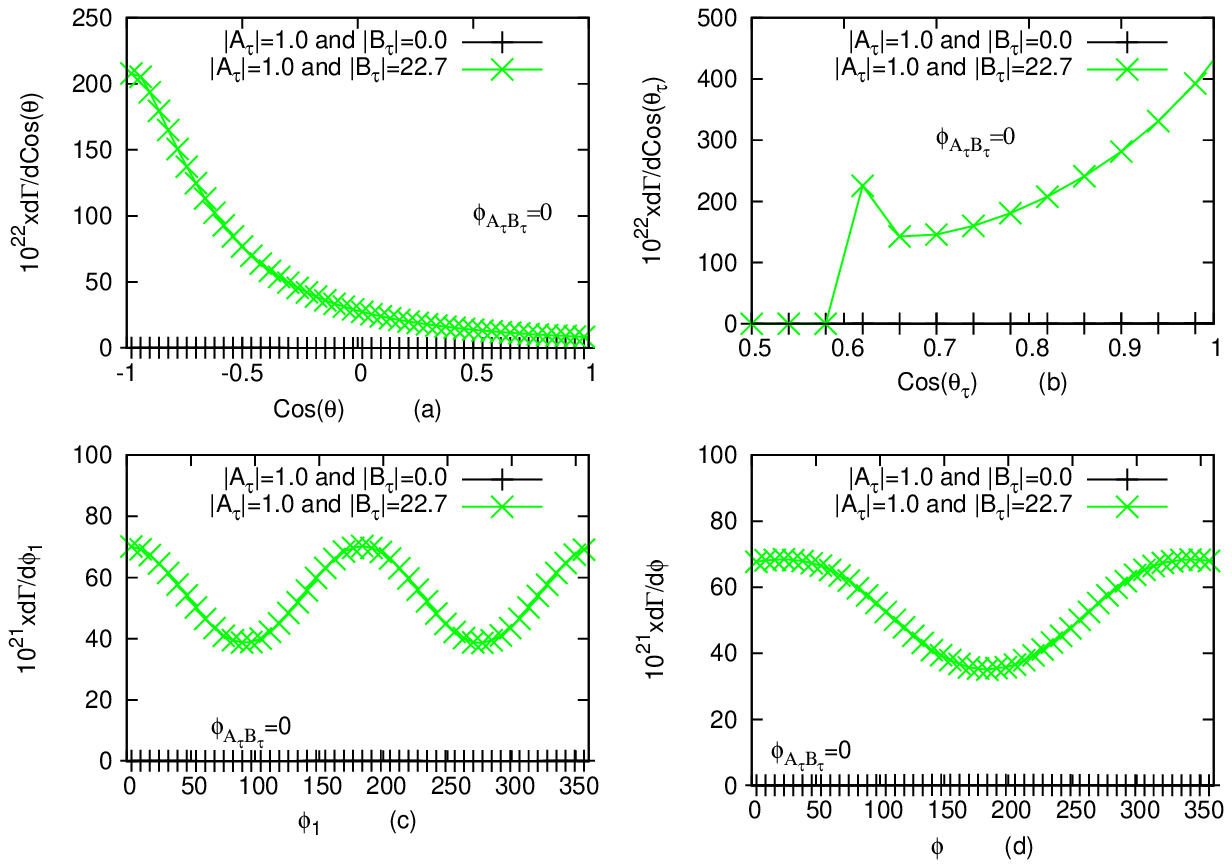}}
\caption{(a) Variation of observable $\frac{d\Gamma}{dCos\theta}$ as a function of $Cos\theta$ for $|A_\tau|=1$ and smallest and largest values of $|B_\tau|$ for Case-VI, VII and VIII.
(b) Variation of observable $\frac{d\Gamma}{dCos\theta_\tau}$ as a function of $Cos\theta_\tau$ for $|A_\tau|=1$ and smallest and largest values of $|B_\tau|$ for Case-VI, VII and VIII.
(c) Variation of observable $\frac{d\Gamma}{d\phi_1}$ as a function of $\phi_1$ for $|A_\tau|=1$ and smallest and largest values of $|B_\tau|$ for Case-VI, VII and VIII.
(d) Variation of observable $\frac{d\Gamma}{d\phi}$ as a function of $\phi$ for $|A_\tau|=1$ and smallest and largest values of $|B_\tau|$ for Case-VI, VII and VIII.
For all these diagrams we have chosen $A_\tau=1$ and the relative phase $\phi_{A_\tau\,B_\tau}$ between $A_\tau$ and $B_\tau$ as zero. }
\label{fig:fig8}
\end{figure}

In previous section we have shown how to detect the presence of NP from the decay width and angular analysis of the cascade decay $B^0_{s}\to \tau^+\tau^- \to (\rho^+\bar{\nu}_\tau)_{\tau^+}(\pi^{-}\nu_{\tau})_{\tau^-} \to ((\pi^+ \pi^0)_{\rho^+}\bar{\nu}_\tau)_{\tau^+}(\pi^{-}\nu_{\tau})_{\tau^-}$. We have explained that if NP is present, how to figure out it's Lorentz structure using our analysis for the mentioned cascade decay. But then next question which comes in mind is ``will we be able to separate the effect of pure NP from the effect of pure SM or a combination of SM+NP?" To answer this question, we notice from Eq.~(\ref{eq:AB}) that $B_\tau$ depends only on $NP_3$($C_{PS}$) whereas $A_{\tau}$ depends on SM($C_{AA}^{SM}$), $NP_1$($C_{AA}^{NP}$) and $NP_2$($C_{PP}$). This signifies that from $A_{\tau}$, we cannot isolate the NP from SM. But if we can establish some difference due to the presence of $B_\tau$ only, then that should be a definite indication of NP. In this section our main aim is to show whether we can isolate $B_\tau$ dependence or not. \\ 

First of all only in the presence of SM, $A_\tau$ is restricted between $19-21.5$ and $B_\tau=0$. On average if we choose $A_\tau=20$, it is allowed by all possible cases of Table.~(\ref{tab:decaywidth}) but $B_\tau=0$ in the absence of $NP_3$. For $A_\tau=20$, $B_\tau$ can be as large as 11.4 for Case-IV whereas for Case-VI, VII and VIII $B_\tau$ can be as large as 22.7. We plot $\frac{d\Gamma}{dcos\theta}$ and $\frac{d\Gamma}{dcos\theta_\tau}$ in Fig.~\ref{fig:fig7}(a) and (b) respectively for $A_\tau=20, \, B_\tau=0$ and $A_\tau=20, \, B_\tau=22.7$. There is a clear difference between these two situations, hence, there is a possibility to isolate $NP_3$ from SM for Case-VI, VII and VIII. For Case-IV we have already shown in Fig.~\ref{fig:fig3} that with precision measurement the presence of $NP_3$ can be isolated. Plot for $\frac{d\Gamma}{d\phi_1}$ and $\frac{d\Gamma}{d\phi}$ in Fig.~\ref{fig:fig7}(c) and (d) also emphasizes on the same conclusion.\\

Further we notice that the partial decay widths are sensitive to $B_\tau$ for small values of $A_\tau$. This can be seen from Fig.~\ref{fig:fig4}. Small value of $A_\tau$ is chosen for Fig.~\ref{fig:fig4}(c)-(d), whereas for Fig.~\ref{fig:fig4}(a)-(b) sufficiently large value of $A_\tau$ is chosen. Presence of non-zero $B_\tau$ is more prominent for Fig.~\ref{fig:fig4}(c)-(d) than Fig.~\ref{fig:fig4}(a)-(b). Using this fact we can further isolate the effect of $B_\tau$. $A_\tau=1$ is allowed for Case-II, III, V as well as for Case-IV, VI, VII and VIII. The main difference is that for Case-II, III, V, $NP_3$ is not present i.e. $B_\tau=0$ for $A_\tau=1$ whereas for Case-VI, VII and VIII, $NP_3$ is present that means $B_\tau$ can be equal to 22.7 for $A_\tau=1$. Plots for partial decay widths $\frac{d\Gamma}{dcos\theta}$, $\frac{d\Gamma}{dcos\theta_\tau}$, $\frac{d\Gamma}{d\phi_1}$ and $\frac{d\Gamma}{d\phi}$ in Fig.~\ref{fig:fig8}(a),(b),(c) and (d) shows a clear difference for these two different situations. This way too we can isolate the effect of pure $NP_3$ dependence.\\

\section{IV. Summary and Conclusion}

In this paper we have computed the bounds on the couplings of the loop mediated general Hamiltonian involved in the leptonic decays of $B^0_s$ meson. We have implemented several features in this analysis which were not considered into account earlier. Present experimental bound on $\mathcal{BR}( B_s^0 \to \mu^+ \mu^-)$ is considered in our analysis (previously we have only the upper bounds). In this study we have worked in a completely model independent fashion and have included the couplings allowed by all possible Lorentz structure in this decay mode. The possibility of both real or complex couplings have been considered. As the experimental and SM estimated branching ratios for $B_s^0 \to \mu^+ \mu^-$ are in the same regime, NP couplings are much more restricted now. Interference between SM and various NP couplings play a very crucial role in determining the bounds, particularly when both $NP_1(C_{AA}^{NP})$ and $NP_2(C_{PP})$ are simultaneously present. There is an intricate interplay among different amplitudes. \\

Similar couplings are going to affect the branching ratio of $B_s^0 \to \tau^+\tau^-$. To observe the effect due to the presence of NP, we have explored the angular analysis of $B^0_s\to\tau^{+}\tau^{-}\to(\rho^+\bar{\nu}_\tau)_{\tau^+}(\pi^{-}\nu_{\tau})_{\tau^-}\to ((\pi^+\pi^0_{\rho^+}+\bar{\nu}_\tau)_{\tau^+}(\pi^{-}\nu_{\tau})_{\tau^-}$ cascade decay. In this angular analysis we have suggested five observables, which can provide sizeable difference from SM in the presence of NP. Not only that, for the simultaneous presence of $NP_1$, $NP_2$, $NP_3$ or $NP_1$, $NP_3$ or $NP_2$,  $NP_3$ we have managed to show an analysis to isolate purely NP effects from SM effects too, which is quite an accomplishment. \\

\section{V. Acknowledgements}
We sincerely thank  Prof. Rahul Sinha and Prof. Anirban Kundu for their extremely valuable comments and suggestions. The work of Jyoti P. Saha is supported by DST-PURSE, Govt. of India.

\section{VI. Appendix}

The partial decay width of the cascade decay $B^0_{s}\to \tau^+\tau^- \to (\rho^+\bar{\nu}_\tau)_{\tau^+}(\pi^{-}\nu_{\tau})_{\tau^-} \to ((\pi^+ \pi^0)_{\rho^+}\bar{\nu}_\tau)_{\tau^+}(\pi^{-}\nu_{\tau})_{\tau^-}$ has been expressed in Eq.~(\ref{eq:d-phase-space}). It involves the matrix element for this process. This matrix element has terms containing $|A_\tau|^2$, $|B_\tau|^2$, $\mathrm{Real}(A_\tau\, B_\tau^*)$ and $\mathrm{Img}(A_\tau\, B_\tau^*)$. The coefficients of $|A_\tau|^2$ has been listed in Table.(III) . The coefficients of $|B_\tau|^2$ has been listed in Table.(IV) whereas the coefficients of $\mathrm{Real}(A_\tau\, B_\tau^*)$ has been listed in Table.(V) and the coefficients of $\mathrm{Img}(A_\tau\, B_\tau^*)$ has been listed in Table. (VI).

\begin{table*}[thb]
  \centering
  \begin{tabular}{|p{0.25cm}l|p{11cm}|}
    \hline
    \hline
    \multicolumn{2}{|l|}{
      $1$}&\\[1.5ex]
    \hline
    \mbox{}&$\times 1$& $\begin{array}{l}m_{\rho }^4 m_{\tau }^4
      \Big(-2 m_{\tau }^6+8 k M^2 q_1\left(M q_1-m_{\rho
        }^2\right)1-8 k M q_1 \left(M+4 q_1\right) m_{\tau
      }^2\\+2\left(-3 m_{\rho }^2+M \left(2
          q_1+\sqrt{q^2+m_{\rho}^2}\right)+k \left(M+8 q_1-4 
          \sqrt{q^2+m_{\rho }^2}\right)\right) m_{\tau
      }^4\\+\left(\left(M\left(M+4 q_1\right)+2 k \left(M+8
            q_1\right)\right) m_{\rho }^2+4 \sqrt{q^2+m_{\rho
          }^2}\left(4 k-M\right) M q_1\right) m_{\tau }^2\Big) 
\end{array}$\\[1.5ex]   
    \hline
    \mbox{}&$\times \cos\theta$&$2 k q m_{\rho }^4 m_{\tau }^4 M^2
    \left(m_{\tau }^2-2 M q_1\right) $\\[1.5ex]  
    \hline
    \mbox{}&$\times \cos\theta\cos\theta_1$&$-4 k m_{\rho }^4 m_{\tau
    }^4 \left(M^2-4 m_{\tau }^2\right) q_1 \left(-m_{\rho
      }^2-m_{\tau }^2+2 M q_1\right)$\\[1.5ex]  
    \hline
    \mbox{}&$\times \cos\theta_1\cos\theta_\tau$& $\begin{array}{l}4 y
      m_{\rho }^4 m_{\tau }^4 M q_1 \Big(-2 m_{\tau }^4+2 k M
      \left(m_{\rho }^2-2 M q_1\right)\\+\left(-2 m_{\rho
        }^2+\sqrt{q^2+m_{\rho }^2} M+8 k q_1+2 M
        \left(k+q_1\right)-4 k \sqrt{q^2+m_{\rho}^2}\right) m_{\tau
      }^2\Big)\end{array}$\\[1.5ex] 
    \hline
    \mbox{}&$\times \cos\theta\cos\theta_\tau$&$-2 k y m_{\rho }^4 m_{\tau }^4 M \left(-m_{\rho }^2-m_{\tau }^2+\sqrt{q^2+m_{\rho }^2} M\right)
   \left(2 M q_1-m_{\tau }^2\right)$\\[1.5ex]    
    \hline
    \mbox{}&$\times \cos\theta\cos\theta_1\cos\theta_\tau$&$-4 k q y m_{\rho }^4 m_{\tau }^4 M \left(M^2-4 m_{\tau }^2\right) q_1$\\[1.5ex] 
    \hline
    \mbox{}&$\times \cos\theta\cos^2\theta_1\cos\theta_\tau$&$8 k y m_{\rho }^4 m_{\tau }^4 M \left(M^2-4 m_{\tau }^2\right) q_1^2$\\[1.5ex]  
    \hline
    \mbox{}&$\times \cos\theta\cos\theta_1\cos^2\theta_\tau$&$4 k y^2 m_{\rho }^4 m_{\tau }^4 M^2 q_1 \left(-2 m_{\rho }^2-2 m_{\tau }^2+M \left(2
   q_1+\sqrt{q^2+m_{\rho }^2}\right)\right)  $\\[1.5ex]
    \hline
    \mbox{}&$\times \cos^2\theta_1\cos^2\theta_\tau$&$8 y^2 m_{\rho }^4 m_{\tau }^4 M^2 \left(k M-m_{\tau }^2\right) q_1^2$\\[1.5ex]
    \hline
    \mbox{}&$\times \cos\theta\cos^2\theta_1\cos^3\theta_\tau$&$-8 k y^3 m_{\rho }^4 m_{\tau }^4 M^3 q_1^2$\\[1.5ex]
    \hline
    \hline
    \multicolumn{2}{|l|}{
      $\sin\theta\,\sin\theta_1$}&\\[1.5ex]
    \hline
    \mbox{}&$\times 1$& $-4 k \cos \left(\phi -\phi _1\right) m_{\rho
    }^4 m_{\tau }^4 \left(4 m_{\tau }^2-M^2\right) q_1\left(m_{\rho
      }^2+m_{\tau }^2-2 M q_1\right)$\\[1.5ex]   
    \hline
    \mbox{}&$\times \cos\theta_1\cos\theta_\tau$&$8 k y \cos
    \left(\phi -\phi _1\right) m_{\rho }^4 m_{\tau }^4 M
    \left(M^2-4 m_{\tau }^2\right) 
   q_1^2 $\\[1.5ex] 
    \hline
    \hline
    \multicolumn{2}{|l|}{
      $\sin\theta_1\,\sin\theta_\tau$}&\\[1.5ex]
    \hline
    \mbox{}&$\times 1$&
    $\begin{array}{l}
      4 y \cos\phi_1 m_{\rho }^4 m_{\tau }^4 M q_1 \Big(-2
      m_{\tau }^4+2 k M   \left(m_{\rho }^2-2 M
        q_1\right)\\+\left(-2 m_{\rho }^2+\sqrt{q^2+m_{\rho }^2}
        M+8 k q_1+2 M \left(k+q_1\right)-4 k
        \sqrt{q^2+m_{\rho}^2}\right) m_{\tau }^2\Big) 
    \end{array}
    $\\  
    \hline
    \mbox{}&$\times \cos\theta$ & $-4 k q y \cos\phi_1
    m_{\rho }^4 m_{\tau }^4 M \left(M^2-4 m_{\tau }^2\right)
    q_1$\\[1.5ex]   
    \hline
    \mbox{}&$\times \cos\theta\cos\theta_1$&$8 k y \cos\phi_1 m_{\rho
    }^4 m_{\tau }^4 M \left(M^2-4 m_{\tau}^2\right) q_1^2$\\[1.5ex]   
    \hline
    \mbox{}&$\times\cos\theta \cos\theta_\tau$&$4 k y^2 \cos\phi_1
    m_{\rho }^4 m_{\tau }^4 M^2 q_1 \left(-2 m_{\rho }^2-2
      m_{\tau}^2+M \left(2 q_1+\sqrt{q^2+m_{\rho 
          }^2}\right)\right)$\\[1.5ex]    
    \hline
    \mbox{}&$\times\cos\theta_1 \cos\theta_\tau$&$16 y^2 \cos\phi_1
    m_{\rho }^4 m_{\tau }^4 M^2 \left(k M-m_{\tau }^2\right)
    q_1^2$\\[1.5ex]     
    \hline
    \mbox{}&$\times\cos\theta\cos\theta_1\cos^2\theta_\tau$&$-16 k y^3
    \cos\phi_1 m_{\rho }^4 m_{\tau }^4 M^3 q_1^2$\\[1.5ex]     
    \hline
    \hline
    \multicolumn{2}{|l|}{
      $\sin\theta\,\sin\theta_\tau$}&\\[1.5ex]
    \hline
    \mbox{}&$\times 1$&$ -2 k y \cos\phi m_{\rho }^4 m_{\tau }^4
    M \left(m_{\rho }^2+m_{\tau }^2-\sqrt{q^2+m_{\rho }^2}
      M\right) \left(m_{\tau }^2-2 M q_1\right)  $\\[1.5ex]   
    \hline
    \mbox{}&$\times \cos\theta_1\cos\theta_\tau$&$ 4 k y^2 \cos\phi
     m_{\rho }^4 m_{\tau }^4 M^2 q_1 \left(-2 m_{\rho }^2-2 m_{\tau
      }^2+M \left(2 q_1+\sqrt{q^2+m_{\rho }^2}\right)\right)
    $\\[1.5ex]    
    \hline
    \mbox{}&$\times \cos^2\theta_1\cos^2\theta_\tau$&$-8 k y^3
    \cos\phi m_{\rho }^4 m_{\tau }^4 M^3 q_1^2 $\\[1.5ex]     
    \hline
    \hline
    \multicolumn{2}{|l|}{
      $\sin\theta\,\sin^2\theta_1\,\sin\theta_\tau$}&\\[1.5ex]
    \hline
    \mbox{}&$\times 1$&$8 k y \cos \left(\phi -\phi _1\right)
    \cos\phi_1 m_{\rho }^4 m_{\tau }^4 M \left(M^2-4 
      m_{\tau }^2\right) q_1^2$\\[1.5ex]   
    \hline
    \hline
    \multicolumn{2}{|l|}{
      $\sin\theta\,\sin\theta_1\,\sin^2\theta_\tau$}&\\[1.5ex]
    \hline
    \mbox{}&$\times 1$&$-16 k y^3 \cos\phi \cos\phi_1 m_{\rho }^4
    m_{\tau }^4 $\\[1.5ex]    
    \hline
    \mbox{}&$\times \cos\theta_1\cos\theta_\tau$&$ -16 k y^3 \cos\phi
    \cos\phi_1 m_{\rho }^4 m_{\tau }^4 M^3 
    q_1^2$\\[1.5ex]     
    \hline
    \hline
    \multicolumn{2}{|l|}{
      $\sin^2\theta_1\,\sin^2\theta_\tau$}&\\[1.5ex]
    \hline
    \mbox{}&$\times 1$&$8 y^2 \cos ^2\phi _1 m_{\rho }^4
    m_{\tau }^4 M^2  \left(k M-m_{\tau }^2\right) q_1^2$\\[1.5ex]     
    \hline
    \mbox{}&$\times \cos\theta\cos\theta_\tau$&$-8 k y^3 \cos\phi
    \cos ^2\phi_1 m_{\rho }^4 m_{\tau }^4  $\\[1.5ex]      
    \hline
    \hline \multicolumn{2}{|l|}{
      $\sin\theta\sin^2\theta_1\,\sin^3\theta_\tau$}&\\[1.5ex]
    \hline
    \mbox{}&$\times 1$&$-8 k y^3 \cos\phi \cos^2\phi_1 m_{\rho }^4
    m_{\tau }^4 M^3 q_1^2$\\[1.5ex] 
    \hline
  \end{tabular}
  \caption{Coefficients of $|A_\tau|^2$ term in  $|\mathcal{M}|^2$ of
    Eq.~(\ref{eq:d-phase-space})} 
  \label{tab:ASq}
\end{table*}

\begin{table*}[thb]
  \centering
  \begin{tabular}{|p{0.25cm}l|p{11cm}|}
    \hline
    \hline
    \multicolumn{2}{|l|}{
      $1$}&\\[1.5ex]
    \hline
    \mbox{}&$\times 1$&$m_{\rho }^4 m_{\tau }^6 M \left(\left(M-4
        q_1\right) m_{\rho }^2+2 m_{\tau }^2 \left(\sqrt{q^2+m_{\rho
          }^2}-2 q_1\right)+4 M q_1 \left(2 q_1-\sqrt{q^2+m_{\rho
     }^2}\right)\right)$\\[1.5ex]  
    \hline
    \mbox{}&$\times \cos\theta$&$ 2 k q m_{\rho }^4 m_{\tau }^4 M^2
    \left(m_{\tau }^2-2 M q_1\right)   $\\[1.5ex]  
    \hline
    \mbox{}&$\times \cos\theta\cos\theta_1$&$ 4 k m_{\rho }^4 m_{\tau
    }^4 M^2 q_1 \left(-m_{\rho }^2-m_{\tau }^2+2 M
      q_1\right)$\\[1.5ex]  
    \hline
    \mbox{}&$\times \cos\theta_1\cos\theta_\tau$&$4 y m_{\rho }^4
    m_{\tau }^6 M^2 \left(\sqrt{q^2+m_{\rho }^2}-2 q_1\right)
    q_1$\\[1.5ex] 
    \hline
    \mbox{}&$\times \cos\theta\cos\theta_\tau$&$2 k y m_{\rho }^4
    m_{\tau }^4 M^2 \left(2 \left(m_{\rho }^2+M
        \left(\sqrt{q^2+m_{\rho }^2}-2 q_1\right)\right) 
   q_1-m_{\tau }^2 \left(\sqrt{q^2+m_{\rho }^2}-2 
     q_1\right)\right)$\\[1.5ex]    
    \hline
    \mbox{}&$\times \cos\theta\cos\theta_1\cos\theta_\tau$&$4 k q y
    m_{\rho }^4 m_{\tau }^4 M^3 q_1$\\[1.5ex] 
    \hline
    \mbox{}&$\times \cos\theta\cos^2\theta_1\cos\theta_\tau$&$8 k y
    m_{\rho }^4 m_{\tau }^4 M^3 q_1^2$\\[1.5ex]  
    \hline
    \mbox{}&$\times \cos\theta\cos\theta_1\cos^2\theta_\tau$&$-4 k y^2
    m_{\rho }^4 m_{\tau }^4 M^3 \left(\sqrt{q^2+m_{\rho }^2}-2
      q_1\right) q_1$\\[1.5ex]
    \hline
    \hline
    \multicolumn{2}{|l|}{
      $\sin\theta\,\sin\theta_1$}&\\[1.5ex]
    \hline
    \mbox{}&$\times 1$& $4 k \cos \left(\phi -\phi _1\right) m_{\rho
    }^4 m_{\tau }^4 M^2 q_1 \left(-m_{\rho }^2-m_{\tau }^2+2 M
      q_1\right)$\\[1.5ex]  
    \hline
    \mbox{}&$\times \cos\theta_1\cos\theta_\tau$&$-8 k y \cos
    \left(\phi -\phi _1\right) m_{\rho }^4 m_{\tau }^4 M^3
    q_1^2$\\[1.5ex] 
    \hline
    \hline
    \multicolumn{2}{|l|}{
      $\sin\theta_1\,\sin\theta_\tau$}&\\[1.5ex]
    \hline
    \mbox{}&$\times 1$&$4 y \cos\phi_1 m_{\rho }^4
    m_{\tau }^6 M^2 \left(\sqrt{q^2+m_{\rho }^2}-2 q_1\right)
    q_1$\\[1.5ex]  
    \hline
    \mbox{}&$\times \cos\theta$&$4 k q y \cos\phi_1
    m_{\rho }^4 m_{\tau }^4 M^3 q_1$\\[1.5ex]  
    \hline
    \mbox{}&$\times \cos\theta\cos\theta_1$&$-8 k y \cos\phi_1 m_{\rho }^4 m_{\tau }^4 M^3 q_1^2$\\[1.5ex]  
    \hline
    \mbox{}&$\times\cos\theta \cos\theta_\tau$&$-4 k y^2 \cos
   \phi_1 m_{\rho }^4 m_{\tau }^4 M^3
    \left(\sqrt{q^2+m_{\rho }^2}-2 q_1\right) q_1$\\[1.5ex]   
    \hline
    \hline
    \multicolumn{2}{|l|}{
      $\sin\theta\,\sin\theta\,\sin\theta_\tau$}&\\[1.5ex]
    \hline
    \mbox{}&$\times 1$&$2 k y \cos\phi m_{\rho }^4 m_{\tau }^4
    M^2 \left(2 \left(m_{\rho }^2+M \left(\sqrt{q^2+m_{\rho}^2}-2
          q_1\right)\right) q_1-m_{\tau }^2 \left(\sqrt{q^2+m_{\rho
          }^2}-2 q_1\right)\right)$\\[1.5ex]  
    \hline
    \mbox{}&$\times \cos\theta_1\cos\theta_\tau$&$-4 k y^2 \cos\phi
     m_{\rho }^4 m_{\tau }^4 M^3 \left(\sqrt{q^2+m_{\rho }^2}-2
      q_1\right) q_1$\\[1.5ex]   
    \hline
    \hline
    \multicolumn{2}{|l|}{
      $\sin\theta\,\sin^2\theta_1\,\sin\theta_\tau$}&\\[1.5ex]
    \hline
    \mbox{}&$\times 1$&$-8 k y \cos \left(\phi -\phi _1\right)
    \cos\phi_1 m_{\rho }^4 m_{\tau }^4 M^3 q_1^2$\\[1.5ex]   
    \hline
    \hline
    \multicolumn{2}{|l|}{
      $\sin\theta\,\sin\theta_1\,\sin^2\theta_\tau$}&\\[1.5ex]
    \hline
    \mbox{}&$\times 1$&$-4 k y^2 \cos\phi \cos\phi_1 m_{\rho }^4 m_{\tau }^4 M^3
    \left(\sqrt{q^2+m_{\rho}^2}-2 q_1\right) q_1 $\\[1.5ex]  
    \hline
  \end{tabular}
  \caption{Coefficients of $|B_\tau|^2$ term in  $|\mathcal{M}|^2$ of Eq.~(\ref{eq:d-phase-space})}
  \label{tab:BSq}
\end{table*}

\begin{table*}[thb]
  \centering
  \begin{tabular}{|p{0.25cm}l|p{11cm}|}
    \hline
    \hline
    \multicolumn{2}{|l|}{
      $1$}&\\[1.5ex]
    \hline
    \mbox{}&$\times 1$&$\begin{array}{l}m_{\rho }^4 m_{\tau }^4 M \Big(m_{\tau }^2
      \left(\left(-2 k+3 M-2 \sqrt{q^2+m_{\rho }^2}\right) 
        m_{\rho }^2+2 M \left(q^2+4 q_1 \left(q_1-\sqrt{q^2+m_{\rho
              }^2}\right)\right)\right)\\-2 k 
      \left(-m_{\rho }^4+M^2 m_{\rho }^2+M^2 \left(q^2+4 q_1
          \left(q_1-\sqrt{q^2+m_{\rho 
              }^2}\right)\right)\right)\Big)\end{array}$\\[1.5ex] 
    \hline
    \mbox{}&$\times \cos\theta_\tau$&$2 q y m_{\rho }^4 m_{\tau }^4 M \left(2 k
      M^2 \left(\sqrt{q^2+m_{\rho }^2}-2 q_1\right)-m_{\rho}^2
      m_{\tau }^2\right)$\\[1.5ex]  
    \hline
    \mbox{}&$\times  \cos\theta_1 \cos\theta_\tau$&$8 k y m_{\rho }^4
    m_{\tau }^4 M^3 q_1 \left(2 q_1-\sqrt{q^2+m_{\rho
        }^2}\right)$\\[1.5ex]   
    \hline
    \mbox{}&$\times  \cos\theta \cos\theta_\tau$&$ 2 k y m_{\rho }^4
    m_{\tau }^4 M \left(-m_{\rho }^4+\left(m_{\tau }^2+M^2\right)
      m_{\rho }^2+M^2 \left(q^2+4 q_1 \left(q_1-\sqrt{q^2+m_{\rho
            }^2}\right)\right)\right)  $\\[1.5ex]  
    \hline
    \mbox{}&$\times  \cos^2\theta_\tau$&$ y^2 m_{\rho }^4
    m_{\tau }^4 M^2 \left(\left(m_{\rho }^2-2 q^2\right) m_{\tau
      }^2-2 k q^2 M\right)  $\\[1.5ex]   
    \hline
    \mbox{}&$\times  \cos\theta_1 \cos^2\theta_\tau$&$ 8 q y^2 m_{\rho
    }^4 m_{\tau }^4 M^2 \left(m_{\tau }^2+k M\right) q_1
    $\\[1.5ex]   
    \hline
    \mbox{}&$\times  \cos\theta \cos^2\theta_\tau$&$-4 k q y^2 m_{\rho
    }^4 m_{\tau }^4 M^3 \left(\sqrt{q^2+m_{\rho }^2}-2 q_1\right)
    $\\[1.5ex]   
    \hline
    \mbox{}&$\times \cos\theta \cos\theta_1 \cos^2\theta_\tau$&$8 k
    y^2 m_{\rho }^4 m_{\tau }^4 M^3 \left(\sqrt{q^2+m_{\rho }^2}-2
      q_1\right) q_1 $\\[1.5ex]   
    \hline
    \mbox{}&$\times  \cos^2\theta_1 \cos^2\theta_\tau$&$-8 y^2 m_{\rho
    }^4 m_{\tau }^4 M^2 \left(m_{\tau }^2+k M\right) q_1^2
    $\\[1.5ex]   
    \hline
    \mbox{}&$\times  \cos\theta \cos^3\theta_\tau$&$2 k q^2 y^3
    m_{\rho }^4 m_{\tau }^4 M^3 $\\[1.5ex]   
    \hline
    \mbox{}&$\times \cos\theta \cos\theta_1 \cos^3\theta_\tau$&$ -8 k
    q y^3 m_{\rho }^4 m_{\tau }^4 M^3 q_1 $\\[1.5ex]   
    \hline
    \mbox{}&$\times \cos\theta \cos^2\theta_1 \cos^3\theta_\tau$&$8 k
    y^3 m_{\rho }^4 m_{\tau }^4 M^3 q_1^2 $\\[1.5ex]   
    \hline
    \hline
    \multicolumn{2}{|l|}{
      $\sin\theta_1\sin\theta_\tau$}&\\[1.5ex]
    \hline
    \mbox{}&$\times 1$&
    $y^2 m_{\rho }^6 m_{\tau }^6 M^2$\\[1.5ex]
    \hline
    \mbox{}&$\times \cos\theta_\tau$&
    $8 q y^2 \cos\phi_1 m_{\rho }^4 m_{\tau }^4 M^2
    \left(m_{\tau }^2+k M\right) q_1$\\[1.5ex] 
    \hline
    \mbox{}&$\times \cos\theta\cos\theta_\tau$&
    $8 k y^2 \cos\phi_1 m_{\rho }^4 m_{\tau }^4 M^3
    \left(\sqrt{q^2+m_{\rho }^2}-2 q_1\right) q_1$\\[1.5ex]  
    \hline
    \mbox{}&$\times \cos\theta_1\cos\theta_\tau$&
    $-16 y^2 \cos\phi_1 m_{\rho }^4 m_{\tau }^4 M^2
    \left(m_{\tau }^2+k M\right) q_1^2$\\[1.5ex]  
    \hline
    \mbox{}&$\times \cos\theta\cos^2\theta_\tau$&
    $-8 k q y^3 \cos\phi_1 m_{\rho }^4 m_{\tau }^4
    M^3 q_1 $\\[1.5ex] 
    \hline
    \mbox{}&$\times \cos\theta \cos\theta_1\cos^2\theta_\tau$&
    $16 k y^3 \cos\phi_1 m_{\rho }^4 m_{\tau }^4 M^3
    q_1^2 $\\[1.5ex] 
    \hline
    \hline
    \multicolumn{2}{|l|}{
      $\sin\theta\sin\theta_\tau$}&\\[1.5ex]
    \hline
    \mbox{}&$\times 1$&
    $2 k y \cos\phi m_{\rho }^4 m_{\tau }^4 M \left(-m_{\rho
      }^4+\left(m_{\tau }^2+M^2\right) 
   m_{\rho }^2+M^2 \left(q^2+4 q_1 \left(q_1-\sqrt{q^2+m_{\rho
         }^2}\right)\right)\right)  $\\[1.5ex] 
    \hline
    \mbox{}&$\times \cos\theta_\tau$&
    $-4 k q y^2 \cos\phi m_{\rho }^4 m_{\tau }^4 M^3
    \left(\sqrt{q^2+m_{\rho }^2}-2 q_1\right)$\\[1.5ex] 
    \hline
    \mbox{}&$\times \cos^2\theta_\tau$&
    $2 k q^2 y^3 \cos\phi m_{\rho }^4 m_{\tau }^4 M^3 $\\[1.5ex] 
    \hline
    \mbox{}&$\times\cos\theta_1 \cos\theta_\tau$&
    $8 k y^2 \cos\phi m_{\rho }^4 m_{\tau }^4 M^3
    \left(\sqrt{q^2+m_{\rho }^2}-2 q_1\right) q_1 $\\[1.5ex]  
    \hline
    \mbox{}&$\times \cos\theta_1 \cos^2\theta_\tau$&
    $-8 k q y^3 \cos\phi m_{\rho }^4 m_{\tau }^4 M^3 q_1 $\\[1.5ex] 
    \hline
    \mbox{}&$\times\cos^2\theta_1  \cos^2\theta_\tau$&
    $8 k y^3 \cos\phi m_{\rho }^4 m_{\tau }^4 M^3 q_1^2$\\[1.5ex] 
    \hline
    \hline
    \multicolumn{2}{|l|}{
      $\sin\theta\sin\theta_1\sin^2\theta_\tau$}&\\[1.5ex]
    \hline
    \mbox{}&$\times 1$&
    $8 k y^2 \cos\phi \cos\phi_1 m_{\rho }^4
    m_{\tau }^4 M^3 \left(\sqrt{q^2+m_{\rho}^2}-2 q_1\right) q_1
    $\\[1.5ex]  
    \hline
    \mbox{}&$\times \cos\theta_\tau$&
    $-8 k q y^3 \cos\phi \cos\phi_1 m_{\rho }^4
    m_{\tau }^4 M^3 q_1 $\\[1.5ex]  
    \hline
    \mbox{}&$\times \cos\theta_1\cos\theta_\tau$&
    $16 k y^3 \cos\phi \cos\phi_1 m_{\rho }^4
    m_{\tau }^4 M^3 q_1^2 $\\[1.5ex]  
    \hline
    \hline
    \multicolumn{2}{|l|}{
      $\sin^2\theta\sin^2\theta_\tau$}&\\[1.5ex]
    \hline
    \mbox{}&$\times 1$&
    $-8 y^2 \cos ^2\left(\phi _1\right) m_{\rho }^4 m_{\tau }^4 M^2
    \left(m_{\tau }^2+k M\right) q_1^2 $\\[1.5ex]  
    \hline
    \mbox{}&$\times \cos\theta\cos\theta_\tau$&
    $8 k y^3 \cos ^2\left(\phi _1\right) m_{\rho }^4 m_{\tau }^4 M^3
    q_1^2 $\\[1.5ex]  
    \hline
    \hline
    \multicolumn{2}{|l|}{
      $\sin\theta\sin^2\theta_1\sin^3\theta_\tau$}&\\[1.5ex]
    \hline
    \mbox{}&$\times 1$&
    $8 k y^3 \cos\phi \cos ^2\left(\phi _1\right) m_{\rho }^4
    m_{\tau }^4 M^3 q_1^2 $\\[1.5ex]  
    \hline
    \hline
  \end{tabular}
  \caption{Coefficients of $\mathrm{Real}(A_\tau\,B_\tau^*)$ term in
    $|\mathcal{M}|^2$ of  Eq.~(\ref{eq:d-phase-space})} 
  \label{tab:ReAB*}
\end{table*}

\begin{table*}[thb]
  \centering
  \begin{tabular}{|p{0.25cm}l|p{11cm}|}
    \hline
    \hline
    \multicolumn{2}{|l|}{
      $\sin\theta\,\sin\theta_1$}&\\[1.5ex]
    \hline
    \mbox{}&$\times \cos\theta_\tau$&$-2 k M^2 q_1 y \sin (\phi
    -\phi_1) m_{\tau }^4 m_{\rho }^4 \left(m_{\rho }^2+8 M 
      q_1-4 M \sqrt{q^2+m_{\rho }^2}-2 q_1
      \sqrt{q^2+m_{\rho}^2}\right)$\\[1.5ex] 
    \hline
    \mbox{}&$\times \cos\theta_1\cos\theta_\tau$&$-4 k M^2 q q_1^2 y
    \sin (\phi -\phi_1) m_{\tau }^4 m_{\rho }^4$ \\[1.5ex]
    \hline
    \mbox{}&$\times \cos^2\theta_\tau$&$-8 k M^3 q q_1 y^2 \sin (\phi -\phi_1)
    m_{\tau }^4 m_{\rho }^4$\\[1.5ex]
    \hline
    \mbox{}&$\times \cos\theta_1\cos^2\theta_\tau$&$16 k M^3 q_1^2 y^2
    \sin(\phi -\phi_1) m_{\tau }^4 m_{\rho }^4$\\[1.5ex]
    \hline
    \hline
    \multicolumn{2}{|l|}{
      $\sin\theta_1\,\sin\theta_\tau$}&\\[1.5ex]
    \hline
    \mbox{}&$\times \cos\theta$&
    $-2 k y \sin\phi _1 m_{\rho }^4 m_{\tau }^4 M^2
    q_1 \left(m_{\rho }^2-4 \sqrt{q^2+m_{\rho }^2} M+8 
      M q_1-2 \sqrt{q^2+m_{\rho }^2} q_1\right)$\\[1.5ex]
    \hline
    \mbox{}&$\times \cos\theta\cos\theta_1$&
    $-4 k q y \sin\phi _1 m_{\rho }^4 m_{\tau }^4 M^2 q_1^2$\\[1.5ex]
    \hline
    \mbox{}&$\times \cos\theta\cos\theta_\tau$&
    $-8 k q y^2 \sin\phi _1 m_{\rho }^4 m_{\tau }^4 M^3 q_1$\\[1.5ex]
    \hline
    \mbox{}&$\times \cos\theta\cos\theta_1\cos\theta_\tau$&
    $16 k y^2 \sin\phi _1 m_{\rho }^4 m_{\tau }^4 M^3 q_1^2$\\[1.5ex]
    \hline
    \hline
    \multicolumn{2}{|l|}{
      $\sin\theta\,\sin\theta_\tau$}&\\[1.5ex]
    \hline
    \mbox{}&$\times 1$&$k q y \sin\phi  m_{\rho }^4 m_{\tau }^4 M^2
    \left(-3 m_{\rho }^2-2 \left(\sqrt{q^2+m_{\rho }^2}-2 q_1\right) 
      \left(q_1-2 M\right)\right)$\\[1.5ex] 
    \hline
    \mbox{}&$\times \cos\theta_1$&$2 k y \sin\phi m_{\rho }^4 m_{\tau }^4
    M^2 q_1 \left(q^2+m_{\rho }^2-4 \sqrt{q^2+m_{\rho }^2} M+8 M
      q_1-2 \sqrt{q^2+m_{\rho }^2} q_1\right)$\\[1.5ex] 
    \hline
    \mbox{}&$\times \cos\theta_\tau$&$-4 k q^2 y^2 \sin\phi m_{\rho }^4
    m_{\tau }^4 M^3$\\[1.5ex]  
    \hline
    \mbox{}&$\times \cos^2\theta_1\cos\theta_\tau$&$-16 k y^2 \sin\phi
    m_{\rho }^4 m_{\tau }^4 M^3 q_1^2$\\[1.5ex]  
    \hline
    \mbox{}&$\times \cos\theta_1\cos\theta_\tau$&$16 k q y^2 \sin\phi m_{\rho
    }^4 m_{\tau }^4 M^3 q_1$\\[1.5ex] 
    \hline
    \hline
    \multicolumn{2}{|l|}{
      $\sin\theta\,\sin^2\theta_1\,\sin\theta_\tau$}&\\[1.5ex]
    \hline
    \mbox{}&$\times 1$&$-4 k q y \sin \phi m_{\rho }^4 m_{\tau }^4
    M^2 q_1^2$\\[1.5ex] 
    \hline
    \mbox{}&$\times \cos\theta_\tau$&$16 k y^2 \cos\phi_1 \sin
    \left(\phi -\phi _1\right) m_{\rho }^4 m_{\tau }^4 M^3
    q_1^2$\\[1.5ex]  
    \hline
    \hline
    \multicolumn{2}{|l|}{
      $\sin\theta\,\sin\theta_1\,\sin^2\theta_\tau$}&\\[1.5ex]
    \hline
    \mbox{}&$\times 1$&$8 k q y^2 \cos\phi _1 \sin\phi m_{\rho }^4 m_{\tau }^4 M^3 q_1$\\[1.5ex] 
    \hline
    \mbox{}&$\times \cos\theta_1$&$-16 k y^2 \cos\phi _1
    \sin\phi m_{\rho }^4 m_{\tau }^4 M^3 q_1^2$\\[1.5ex]   
    \hline
    \hline
    \multicolumn{2}{|l|}{
      $\sin^2\theta_1\,\sin^2\theta_\tau$}&\\[1.5ex]
    \hline
    \mbox{}&$\times \cos\theta$&$16 k y^2 \cos\phi_1
    \sin\phi_1 m_{\rho }^4 m_{\tau}^4 M^3
    q_1^2$\\[1.5ex]    
    \hline
  \end{tabular}
  \caption{Coefficients of $\mathrm{Img}(A_\tau\,B_\tau^*)$ term in
    $|\mathcal{M}|^2$ of Eq.~(\ref{eq:d-phase-space})}
  \label{tab:ImAB*}
\end{table*}

\end{document}